\PassOptionsToPackage{english}{babel}
\documentclass[aps,showpacs,amssymb,amsfonts,superscriptaddress,twocolumn,pra]{revtex4-1}  %
\usepackage[english,ngerman]{babel}
\usepackage{graphicx}  %
\usepackage{dcolumn}   %
\usepackage{bm}        %
\usepackage{bbm}
\usepackage{amssymb}   %
\usepackage{amsmath}
\usepackage{mathtools}

\usepackage[caption=false]{subfig}

\usepackage{braket}
\usepackage[utf8]{inputenc}
\usepackage{calrsfs}
\DeclareMathAlphabet{\pazocal}{OMS}{zplm}{m}{n}

\usepackage{changes}

\definecolor{myurlcolor}{rgb}{0,0,0.7}
\definecolor{myrefcolor}{rgb}{0.8,0,0}
\usepackage{hyperref}
\hypersetup{colorlinks, linkcolor=myrefcolor,
citecolor=myurlcolor, urlcolor=myurlcolor}

\usepackage{comment}
\usepackage{amsthm}

\newcommand{\ignore}[1]{}

 \definecolor{jordi}{rgb}{0.1,0.1,0.5}
 \newcommand{\jordi}[1]{{\color{jordi} #1}}

\usepackage{times}

\usepackage[capitalize]{cleveref} %
\crefrangeformat{equation}{#3(#1 #4--#5 #2)#6}

\begin{document}
\selectlanguage{english}

\title{Optimization of device-independent witnesses of entanglement depth from two-body correlators}

\author{J. Tura} \email{jordi.tura@mpq.mpg.de} 
\affiliation{Max-Planck-Institut f\"ur Quantenoptik, Hans-Kopfermann-Stra{\ss}e 1, 85748 Garching, Germany}

\author{A. Aloy}
\affiliation{ICFO - Institut de Ciencies Fotoniques, The Barcelona Institute of Science and Technology, 08860 Castelldefels (Barcelona), Spain}

\author{F. Baccari}
\affiliation{ICFO - Institut de Ciencies Fotoniques, The Barcelona Institute of Science and Technology, 08860 Castelldefels (Barcelona), Spain}

\author{A. Acín}
\affiliation{ICFO - Institut de Ciencies Fotoniques, The Barcelona Institute of Science and Technology, 08860 Castelldefels (Barcelona), Spain}
\affiliation{ICREA, Pg. Lluis Companys 23, 08010 Barcelona, Spain}

\author{M. Lewenstein}
\affiliation{ICFO - Institut de Ciencies Fotoniques, The Barcelona Institute of Science and Technology, 08860 Castelldefels (Barcelona), Spain}
\affiliation{ICREA, Pg. Lluis Companys 23, 08010 Barcelona, Spain}

\author{R. Augusiak}
\affiliation{Center for Theoretical Physics, Polish Academy of Sciences, Aleja Lotnik\'ow 32/46, 02-668 Warsaw, Poland}

\date{\today}

\begin{abstract}
In a recent work [A. Aloy \textit{et al.}, \href{https://arxiv.org/abs/1807.06027}{arXiv:1807:06027} (2018)] we have considered the characterization of entanglement depth, from a device-independent perspective, in a quantum many-body system. We have shown that the inequalities introduced in [J. Tura \textit{et al.}, \href{https://doi.org/10.1126/science.1247715}{Science {\bf 344} 1256} (2014)] can be used to obtain device-independent witnesses of entanglement depth and that they enjoy two key properties that allow to compute their $k$-producibility bounds more efficiently for larger system sizes, as well as yielding experimentally-friendlier device-independent witnesses of entanglement depth: they involve at most two-body correlators and they are permutationally invariant.
While the main aim of our previous work was to illustrate the main ideas and applicability of the method, here we outline the details and complement its findings with detailed analysis and further case studies. Specifically, we consider the problem of finding the $k$-producible bounds of such DIWEDs under different assumptions. Not surprisingly, with the weakest assumptions, we can compute $k$-producible bounds only for relatively small number of parties; however we can still learn interesting features from these solutions that motivate the search on larger systems under the assumption that these features persist. This allows us to tackle the case where the system size eventually reaches the thermodynamic limit.

\end{abstract}

\maketitle

\section{Introduction}

Assessing entanglement in many-body systems is a fundamental question that requires dedicated tools \cite{AmicoRMP2008}. On the one hand, full tomography on the state is impractical. On the other hand, a practical criterion should be as easy to measure as possible; for instance, it is desirable that such a criterion consists of few-body correlators (where \textit{few} should ideally be \textit{two}). Furthermore, the lower the number of assumptions on the criterion being used, the better. With the advent of the device-independent (DI) formalism for quantum information processing \cite{AcinDIQKD}, entanglement witnesses have been naturally upgraded to their DI version \cite{BancalPRL2011a, YCLiangEntDepth, MoroderPRL2013, Lin2019}.
Recently, we have considered the characterization of entanglement depth in a many-body system from a device-independent perspective \cite{Aloy2018}. Our criteria are based on the families of Bell inequalities introduced in \cite{SciencePaper}, yielding device-independent witnesses of entanglement depth (DIWEDs). Such DIWEDs naturally inherit desirable properties that allow (i) to compute their $k$-producibility bounds more efficiently for larger system sizes and (ii) they are experimentally friendlier to measure, as they involve at most two-body correlators and are permutationally invariant.

DIWEDs are stronger than usual entanglement witnesses, in the sense that they are derived without relying on assumptions about the physical systems (such as the dimension of the underlying Hilbert space) nor the measurements that are performed on them \cite{BancalPRL2011a}. In addition, in the many-body regime, where a loophole-free Bell test may be very difficult to achieve in practice, the DIWEDs we consider in this work can also be indirectly measured via trusted collective measurements and second moments thereof, opening the way to probe them in situations beyond the current state-of-the-art \cite{SchmiedScience2016, EngelsenPRL2017}.

Determining how many particles share genuinely multipartite entanglement constitutes a good and intuitive measure of the entanglement strength present in the system. While in \cite{Aloy2018} our aim was to illustrate the key ideas and the applicability of our method, here we analyze the construction of such DIWEDs in detail, providing further case studies. We discuss the different methods to derive $k$-producible bounds under different sets of assumptions. Interestingly, some of the features that we observe in the smallest systems (where one can make the weakest assumptions) motivate the search on larger system sizes under the assumption that these features persist, eventually reaching the thermodynamic limit. In this work, we also compare the efficiency of our DIWEDs against other witnesses of entanglement depth, witnesses of nonlocality depth and actual experimental data.

The paper is organized as follows: In Section \ref{sec:prelim} we set the notation and recall the basic concepts this work is built upon. In Section \ref{sec:kprod} we discuss the optimization problem for finding $k$-producibility bounds in a many-body systems under different families of Bell inequalities, from more general to the case where they are permutationally invariant and consisting of one- and two-body correlators only. In Section \ref{sec:methods} we describe in detail the two complementary methods that we use to characterize the $k$-producibility bounds on the DIWEDs we present: a variational method and a certificate of optimality based on a partial solution of the quantum marginal problem.
In Section \ref{sec:numerical} we present two case studies and the numerical results that stem from our work; the first one without extra assumptions and the second one assuming the features that numerics suggest.
In Section \ref{sec:asymptotic} we perform an asymptotic analysis that allows us to delve fully into the many-body regime, obtaining the scaling of different $k$-producibility bounds, where $k$ is a function of the system size. In Section \ref{sec:experimental} we discuss how, contrary to existing DIWEDs, the ones we present here can be accessed through experimentally-friendly observables, such as collective measurements and moments thereof. In Section \ref{sec:otherent} we compare the entanglement depth criteria we derive in this work with previous existing criteria. Finally, we conclude and discuss further research directions in Section \ref{sec:concl}.

\section{Preliminaries} \label{sec:prelim}
We begin this section by introducing the concepts that will be used throughout all the paper. We review the relevant separability definitions and some natural extension to the multipartite case in Section \ref{subsec:sep}. In Section \ref{subsec:BellNonlocalityEntanglementDetection} we review the use of Bell nonlocality for entanglement detection. In particular, in Section \ref{subsubsec:BellExp} we recall the definition of a Bell experiment in the multipartite case, and then in Section \ref{subsubsec:BI} we provide the basics of Bell inequalities involving two-body correlators.

\subsection{Separability notions}\label{subsec:sep}
Let us begin by considering a bipartite quantum system shared by Alice (A) and Bob (B), which is described by a bipartite Hilbert space ${\pazocal H} = {\pazocal H}_A \otimes {\pazocal H}_B$. Let us denote by ${\cal B}({\pazocal H})$ the set of bounded linear operators acting on $\pazocal H$. Then, a mixed state, represented by a density matrix $\rho \in {\cal B}({\pazocal H})$, is a positive-semidefinite operator ($\rho \succcurlyeq 0$) of unit trace, $\mathrm{Tr}(\rho) = 1$. A mixed state $\rho$ is said to be separable if it cannot be written in the following form \cite{WernerPRA1989}:
\begin{equation}
 \label{eq:def:separable}
 \rho = \sum_{i} \lambda_i \rho_i^{(A)} \otimes \rho_i^{(B)},
\end{equation}
where $\lambda_i$ form a convex combination (they are non-negative numbers that sum $1$). \cref{eq:def:separable} has the following operational interpretation: Alice and Bob, in distant laboratories, can prepare a separable state $\rho$ from scratch just by pre-agreeing on a common strategy (given by the $\lambda_i$'s) and preparing locally the corresponding states $\rho_i^{(A)}$ ($\rho_i^{(B)}$) \cite{WernerPRA1989, ChitambarCMP2014}.

\begin{figure}[h]
\centering
\subfloat[Fully separable, $1$-producible]{\includegraphics[width=0.18\textwidth]{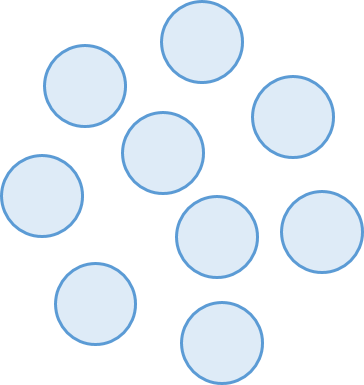}}
\qquad\quad
\subfloat[GME, $9$-producible]{\includegraphics[width=0.18\textwidth]{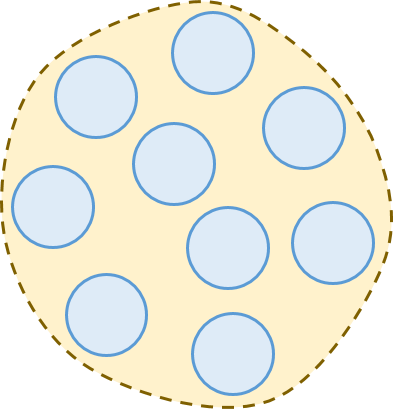}} \\
\subfloat[$4$-separable, $3$-producible]{\includegraphics[width=0.18\textwidth]{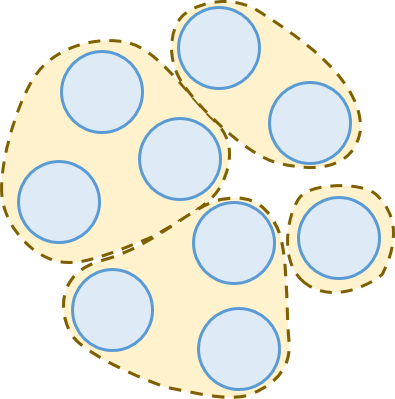}}
\qquad\quad
\subfloat[$4$-separable, $5$-producible]{\includegraphics[width=0.18\textwidth]{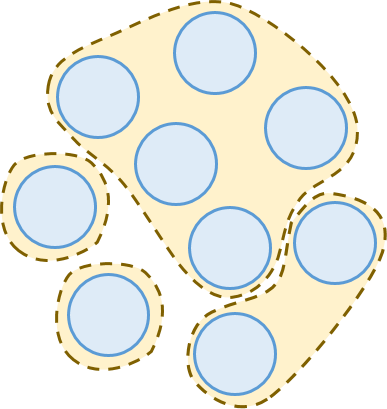}}
\caption{Cartoon picture illustrating the notions of $k$-producibility and $l$-separability. Operationally, the parties (represented by blue circles) in different groups (represented by shapes enclosing a subset of the parties) are allowed to perform local operations and classical communication between different groups. Within each group, different parties can prepare any state they want. Note that we consider a framework where, at every round, groupings may change (cf. \cref{eq:def:kseparable} and \cref{eq:def:kprod}), albeit maintaining the restrictions imposed by $k$ and $l$. Note that $k$ limits the maximal number of parties in any group, while $l$ limits the total number of groups.}
  \label{fig:sepnotions}
\end{figure}

In the multipartite setting, the notion of separability admits many generalizations \cite{SzalayPRA2012, SzalayPRA2015}. A natural one is the so-called biseparability. Let us consider a set of $n$ parties, now indexed by $[n]:=\{0, \ldots, n-1\}$. Following the above interpretation, a biseparable state is defined as a state that can be produced by allowing the parties to join in two (non-empty) subsets and produce the state in the same fashion. Moreover, the parties are allowed to join in different groups over different realizations, as long as they never join all together. Then, the density matrix that describes the quantum state produced by this procedure is given by
\begin{equation}
 \label{eq:def:biseparable}
 \rho = \sum_{A \subset[n], A \neq [n], \emptyset} \lambda_A \sum_{i} \lambda_i^{A} \rho_i^{(A)}\otimes \rho_i^{(A^c)},
\end{equation}
where $A^c := [n] \setminus A$ is the complement of $A$ in $[n]$, and the $\lambda_A$'s form convex combinations. This notion can be immediately generalized (see e.g. \cite{SzalayPRA2012, LeviPRL2013}) to that of $k$-separability by simply considering the possible partitions of $[n]$ into $k$ pairwise disjoint, nonempty subsets, i.e., $A^1, \ldots, A^k \subsetneq [n]$ such that $A^i\cap A^j = \emptyset$ if $i \neq j$, and $\cup_{i=1}^{k} A^i = [n]$ and $A^i \neq \emptyset$. Let us denote by ${\cal P}^k_{\mathrm{sep}}$ the set of partitions of $[n]$ into $k$ subsets satisfying such conditions. Then, a $k$-separable state is described by a density matrix that admits the following decomposition:
\begin{equation}
 \label{eq:def:kseparable}
 \rho = \sum_{{\pazocal P} \in {\mathcal P}^k_{\mathrm{sep}}} \lambda_{\pazocal P} \sum_{i} \lambda_i^{\pazocal P} \bigotimes_{A^j \in {\pazocal P}}\rho_i^{(A^j)}.
\end{equation}
As we shall argue later on in the paper, in \cref{sec:numerical}, the definition of biseparability introduced in \cref{eq:def:biseparable} (or, more generally, the definition of $k$-separability in \cref{eq:def:kseparable}), although being a natural generalization of \cref{eq:def:separable}, are not ideal from an experimental point of view. In fact, the notion of $k$-producibility is much better suited to our work, which we define in what follows:

If, instead of restricting the size of the partitions of $[n]$ to consist of $k$ proper nonempty subsets, we restrict the size of the elements in the partition to contain, at most, $k$ elements, we obtain the definition of $k$-producibility (see e.g. \cite{GuehneNJP2010, WoelkNJP2016}). Let $|S|$ denote the number of elements in a set $S$. Let ${\mathcal P}^k_{\mathrm{prod}}$ be the set of partitions ${\pazocal P}$ of $[n]$ satisfying $A_i \subsetneq[n]$, $A_i \cap A_j = \emptyset$ if $i \neq j$, $|A_i| \leq k$ and $\cup_{A_i \in {\pazocal P}} A_i = [n]$ for all $A_i \in {\pazocal P} $.
Then, a multipartite quantum state is said to be $k$-producible if, and only if, it admits the following decomposition:
\begin{equation}
 \label{eq:def:kprod}
 \rho = \sum_{{\pazocal P} \in {\mathcal P}^k_{\mathrm{prod}}} \lambda_{\pazocal P} \sum_{i} \lambda_i^{\pazocal P} \bigotimes_{A_j \in {\pazocal P}} \rho_i^{(A_j)}.
\end{equation}
A quantum state that cannot be written in the form of \cref{eq:def:kprod} is said to have an entanglement depth at least $k+1$.
Operationally, the interpretation is that strictly more than $k$ parties would need to join together, in order to prepare such a state from scratch. We shall denote by ${\cal D}_k$ the set of $k$-producible states (cf. \cref{eq:def:kprod}).
This definition is clearly preferrable as $k$ indicates the maximal degree of genuinely multipartite entanglement that the parties may be sharing.
In \cref{fig:sepnotions} we schematically outline the differences between such separability notions.

We note that the strongest form of multipartite entanglement, Genuinely Multipartite Entanglement (GME), is given by those states that are not $(n-1)$-producible or, equivalently, not biseparable. Exemplary quantum states that are GME include the GHZ state \cite{Greenberger2007, PanNature2000} and all entangled symmetric states \cite{TuraPRA2012, AugusiakPRA2012, HuberPRA2011}.

\subsection{Bell Nonlocality for entanglement detection}
\label{subsec:BellNonlocalityEntanglementDetection}
It is widely known that every state producing nonlocal correlations must be entangled \cite{WernerPRA1989}. Therefore, Bell inequalities constitute the natural tool to certify entanglement in a device-independent way. Much work has been devoted to their re-interpretation as device-independent entanglement witnesses \cite{YCLiangEntDepth, BancalPRL2011a, MoroderPRL2013} (see also \cite{Lin2019} for a recent complementary approach). On one hand, Bell inequalities can sometimes identify the state producing their maximal violation, thus self-testing the state solely from the statistics it produces in a Bell experiment \cite{MayersYaoST2004, ColadangeloNatComms2017, KaniewskiPRA2017, BampsPRA2015, Bowles2018a, KaniewskiPRL2016, ZhangNPJ2019, BASTA, YangPRAR2013, SATWAP, WangScience2018, Kaniewski2018, ZhangNPJ2019}. Therefore, if one manages to self-test a quantum state of a certain entanglement depth, the goal we pursue in this work is met. However, that only works for very specific states that are known before-hand, as such tests are designed exploiting special properties of the state one wants to self-test.

On the other hand, numerous works have been performed that certify the degree of entanglement present in the state \cite{SchwemmerPRL2015a, KnipsPRL2016, TranPRA2017}, often genuine multipartite entanglement \cite{BancalPRL2011a, Zwerger2018, BarreiroNatPhys2013}. However, the question of assessing the entanglement depth of a many-body system in a device-independent manner remains broadly unexplored, although remarkable steps have been made towards this direction \cite{MoroderPRL2013, YCLiangEntDepth}, yet still focused on few-body systems.

A natural approach to tackle this question for larger values of the entanglement depth and system sizes
consists in designing Bell inequalities for nonlocality depth \cite{Baccari2018}: broadly speaking, one considers hybrid local-nonlocal models of a form similar to \cref{eq:def:kseparable}, where parties in the same group are allowed to produce any no-signalling (possibly supra-quantum) probability distribution. Analogously to the fact that a quantum state violating an ordinary Bell inequality cannot be fully separable, one exploits the property that a state violating such a nonlocality depth Bell inequality (of depth, say, $k$) cannot be $k$-producible.

In this work, however, we consider the case where the parties that form the same group can only produce quantum correlations. This difference is conceptually significant, and it also poses new technical challenges: the set of quantum correlations is much less characterized than its no-signalling counterpart. For instance, the set of no-signalling correlations forms a polyope and it is therefore characterizable by a finite number of linear inequalities, whereas the set of quantum correlations is not even closed \cite{Slofstra2017} and it has a geometrically much richer --albeit complicated-- structure \cite{GohPRA2018}. Despite the increased technical difficulty, constraining the strength of the correlations produced by the parties to be quantum correlations yields DIWEDs which are much tighter than those considered in \cite{Baccari2018}, as we will show in \cref{fig:comparison}.

\subsubsection{The Bell experiment}
\label{subsubsec:BellExp}
Due to the fact that Bell-like inequalities constitute a key element in the derivation of
our DIWEDs, in this section we recall the notion of Bell experiment.
It involves $n$ spatially separated parties sharing a multipartite resource (e.g. a multipartite entangled state). Each party has $m$ measurements at their disposal and each measurement can yield $d$ different outcomes. Let us denote by $\mathbf{a} = (a_0, \ldots, a_{n-1})$ and $\mathbf{x}=(x_0, \ldots, x_{n-1})$ their outputs (outcomes) and inputs (choices of measurements), respectively.
By performing these measurements many times, the probability distributions $p(\mathbf{a}|\mathbf{x})$ are estimated from the collected statistics.
 The parties are allowed to communicate prior to the experiment, but during every round they cannot know any measurement choice from the rest. Also, the parties do not make any assumptions about the internal working of their devices and only the statistics describing the outcomes, given the inputs, are relevant. We denote such a Bell scenario by $(n,m,d)$.

 The no-signalling principle guarantees that the marginal probabilities do not depend on the input of other parties (therefore preventing signalling of information by changing the measurement choice). In particular, in this work, we shall consider only the one- and two-body marginals at sites $i$ and $(i, j)$, respectively, given by
\begin{equation}
 \label{eq:def:12body}
 p_i(a_i|x_i), \qquad p_{ij}(a_i a_j |x_i x_j).
\end{equation}

Of special interest is the multipartite Bell scenario for which $d=2$. In this case, we can assume that the outcomes are labelled $\pm 1$ without loss of generality. Let us denote by ${M}_k^{(i)}$ the correlator corresponding to expectation value of the $k$-th observable of the $i$-th party ($k \in [m]$, $i \in [n]$), which is defined as
\begin{equation}
 M_k^{(i)} = p_i(1|k) - p_i(-1|k).
\end{equation}
Similarly, we consider the two-body correlator corresponding to the $k$-th and $l$-th observables of the $i$-th and $j$-th parties, respectively, which is given by
\begin{equation}
 M_{kl}^{(i,j)} = p_{ij}(a_ia_j=1|kl) - p_{ij}(a_ia_j=-1|kl).
\end{equation}

The notion of correlator can be generalized to any number $p\leq n$ of parties, yielding the $p$-body correlators $M_{k_1,\ldots, k_p}^{i_1,\ldots i_p}$, where $k_j \in [m]$ and $i_j \in [n]$, thus yielding the generic form of a Bell inequality $I - \beta_C \geq 0$, where
\begin{equation}
I:=\sum_{p=1}^n \sum_{k_j \in [m]} \sum_{1 \leq i_1 < \ldots < i_p \leq n} \alpha_{k_1,\ldots, k_p}^{(i_1, \ldots, i_p)}M_{k_1,\ldots, k_p}^{(i_1, \ldots, i_p)},
\label{eq:def:GenericBI}
\end{equation}
with $\alpha_{k_1,\ldots, k_p}^{(i_1, \ldots, i_p)} \in \mathbbm{R}$ and $\beta_C = \min_{\textrm{LHV}}I$ is the so-called classical bound, i.e., the minimum value that $I$ can take over any local hidden variable (LHV) model. Note that this is equivalent to the correlations that arise from $1$-producible states; i.e., product states.

\subsubsection{Bell Nonlocality from symmetric two-body correlators} \label{subsubsec:BI}

We can now consider the following symmetric one- and two-body correlators, respectively defined as \cite{SciencePaper}:
\begin{equation}
 \label{eq:def:symcorr}
 {\pazocal S}_k := \sum_{i\in [n]} {M}_k^{(i)}, \qquad {\pazocal S}_{kl} := \sum_{i \neq j\in [n]} {M}_{kl}^{(i,j)}.
\end{equation}
In this work, we focus on Bell inequalities involving only the correlators introduced in \cref{eq:def:symcorr}. These inequalities have recently appeared in numerous contexts \cite{SciencePaper, AnnPhys, SchmiedScience2016, EngelsenPRL2017, PelissonPRA2016, WagnerPRL2017, FadelPRL2017, Baccari2018, DengPRL2018, FadelQuantum2018, Aloy2018}, and they can be written as 
\begin{equation}
 \label{eq:def:PIBI}
 I:=\sum_{k \in [m]} \alpha_k {\pazocal S}_k + \sum_{k \leq l \in [m]} \alpha_{kl} {\pazocal S}_{kl},
\end{equation}
with $\alpha_k, \alpha_{kl} \in \mathbbm{R}$ and that satisfy $I - \beta_C \geq 0$ on any local hidden variable theory, where $\beta_C$ is the so-called classical bound.

By measuring $I$, one can extract information about the existence of Bell correlations in the system \cite{SciencePaper, AnnPhys, FadelPRL2017} (as has been performed in recent experiments \cite{SchmiedScience2016, EngelsenPRL2017}), even on their nonlocality depth \cite{Baccari2018}.
Therefore, we aim at drawing conclusions on the entanglement depth of the system, given the value of $I$ that is measured. To this aim, we want to find a sequence of values $\beta_k$ such that
\begin{equation}
 \label{eq:def:DIWED}
 I - \beta_k \geq 0
\end{equation}
is satisfied for every possible measurement on any $k$-producible state. We then say that \cref{eq:def:DIWED} constitutes a DIWED.

\section{Optimization over $k$-producible states} \label{sec:kprod}

In this section, we present a general method to find $\beta_k$.
In \cref{subsec:OptGeneric} we describe the method it in its full generality, which applies to any Bell inequality in the $(n,2,2)$ scenario. However, the method can be further improved under certain additional hypotheses: for instance, if the inequality consists of few-body correlators or it enjoys additional symmetries. Therefore, in \cref{subsec:OptTwoBody} we particularize it for the case of two-body correlators. In \cref{subsec:OptTwoBodySymmetric} we add the permutation invariance property and, motivated by the numerical results we shall present in \cref{sec:numerical}, in \cref{subsubsec:OptSymm} we include the additional properties suggested by numerical results.

\subsection{Generic case}
\label{subsec:OptGeneric}
Consider an $(n,2,2)$ scenario and a Bell inequality $I$ of the form of \cref{eq:def:GenericBI}. We begin by noting that, to every Bell inequality $I$, one can assign a Bell operator ${\pazocal B} \in {\cal B}({\pazocal H})$, such that
\begin{equation}
{\pazocal B}:= \sum_{p=1}^n \sum_{k_j \in [m]}\sum_{1 \leq i_1<\ldots < i_p \leq n} \alpha_{k_1,\ldots, k_p}^{(i_1,\ldots,i_p)} {\pazocal M}_{k_1, \ldots, k_p}^{(i_1,\ldots, i_p)},
\label{eq:def:BellOp}
\end{equation}
where
\begin{equation}
{\pazocal M}_{k_1, \ldots, k_p}^{(i_1,\ldots, i_p)} = \bigotimes_{t=1}^p {\pazocal M}_{k_t}^{(i_t)},
\label{eq:def:CorrOp}
\end{equation}
and ${\pazocal M}_k^{(i)} \in {\cal B}({\pazocal H}_i)$ denotes the $k$-th observable performed by the $i$-th party.
Note that with a Bell operator defined as \cref{eq:def:BellOp}, the expectation value of $I$ for a quantum state $\rho$ and local measurements ${\pazocal M}_k^{(i)}$ is automatically given by $I = \mathrm{Tr}[\rho {\pazocal B}]$.

However, observe that the dimension of the local Hilbert spaces ${\pazocal H}_i$ is not fixed. Therefore, in general, one would need to carry the optimization for $\beta_k$ for Hilbert spaces of increasing dimension (see e.g. \cite{PalPRA2010}).
This poses, in general, a formidable challenge, not only from the numerical point of view \cite{JungeJMP2011, FritzRevMathPhys2012}, as an efficient description of the set of quantum correlations is unlikely to be found (see e.g. \cite{NavascuesNatComms2015, NavascuesPRL2007}). It is even undecidable to determine whether a linear-system game admits a perfect finite-dimensional quantum strategy or a limit of finite-dimensional quantum strategies \cite{Slofstra2017}.

Fortunately, in the $(n,2,2)$ case every Bell operator is isometrically equivalent to a Bell operator for qubits (${\cal H}_i = {\mathbbm C}^2$) and, furthermore, one can assume, without loss of generality that the local measurements ${\pazocal M}_k^{(i)}$ are performed on the XZ plane on the Bloch Sphere \cite{Toner2006}
 due to Jordan's lemma \cite{Jordan1875}. Hence, every local measurement can be parametrized as ${\pazocal M}_k^{(i)} = \cos(\theta_{i,k}) \sigma_x^{(i)} + \sin(\theta_{i,k}) \sigma_z^{(i)}$. We shall encode as $\boldsymbol{\theta}$ the vector containing all $\theta_{i,k}$ and we shall therefore denote the Bell operator ${\pazocal B}$ as ${\pazocal B}(\boldsymbol{\theta})$ when we want to make this dependence explicit.

Let us therefore consider $\rho \in {\cal D}_k$, with ${\pazocal H}= ({\mathbbm C}^2)^{\otimes n}$. The optimization of $\beta_k$ is such that
\begin{equation}
\beta_k = \min_{\rho \in {\cal D}_k, \boldsymbol{\theta}} \mathrm{Tr}[\rho {\pazocal B(\boldsymbol{\theta})}].
\end{equation}
Observe that, for every value of $\boldsymbol{\theta}$, one can assume, without loss of generality that $\beta_k(\boldsymbol{\theta})$ is attained at a pure state, which corresponds to a fixed partition ${\pazocal P} \in {\cal P}^k_{\mathrm{prod}}$ (cf. \cref{eq:def:kprod}). Therefore, the optimization of $\beta_k$ can be carried over pure states by fixing their $k$-partition:
\begin{equation}
\beta_k = \min_{\pazocal P \in {\cal P}^k_{\mathrm{prod}}} \beta_k^{\pazocal P},
\end{equation}
where
\begin{equation}
\beta_k^{\pazocal P} = \min_{\ket{\Psi_{\pazocal P}}, \boldsymbol{\theta}} \bra{\Psi_{\pazocal P}} {\pazocal B}(\boldsymbol{\theta}) \ket{\Psi_{\pazocal P}},
\label{eq:partitionfixed}
\end{equation}
and $\ket{\Psi_{\pazocal P}} = \bigotimes_{A \in {\pazocal P}} \ket{\psi_A}$.

In what follows, we show that it is, in principle, possible to find the solutions to the optimization problem \cref{eq:partitionfixed} exactly. This implies tha we can find $\beta_k$ without running into local minima issues. To do that, we use arguments that apply to systems of polynomial equations. In particular, for every $k$-partition $\pazocal P$, there is a system of polynomial equations under polynomial equality constraints that defines $\beta_k^{\pazocal P}$.

To derive such a system, let us define the variables $x_{i,k} \equiv \cos(\theta_{i,k})$ and $y_{i,k} \equiv \sin(\theta_{i,k})$. Clearly, they are subject to the constraint $x_{i,k}^2 + y_{i,k}^2 = 1$ for every $i\in [n]$ and $k \in [2]$. On the other hand, the quantum state ${\ket{\psi_{\pazocal P}}}$ can be assumed to be real, because ${\pazocal B}$ is a real, symmetric operator
. Therefore, we can expand $\ket{\psi_A} = (\psi^A_{0}, \ldots, \psi^A_{2^{|A|}-1})^T$ for every $A \in {\pazocal P}$, where $\sum_{i=0}^{2^{|A|}-1} [\psi_i^{A}]^2 = 1$.

Therefore, $\bra{\Psi_{\pazocal P}} {\pazocal B} \ket{\Psi_{\pazocal P}}$ is a polynomial of $4n+\sum_{A\in {\pazocal P}}2^{|A|}$ variables, of degree $2$ in the $\psi^A_i$ variables and degree at most $n$ in the $x_{i,k}$ and $y_{i,k}$ variables, subject to $2n+|{\pazocal P}|$ equality constraints. One can determine the critical points of $\bra{\Psi_{\pazocal P}} {\pazocal B} \ket{\Psi_{\pazocal P}}$ subject to the above constraints using the Lagrange multipliers method. Defining $g_{i,k} = x_{i,k}^2+y_{i,k}^2-1$ and $h_{A} = \sum_{i=0}^{2^{|A|}-1} [\psi_i^{A}]^2 - 1$ and associating the dual variables $\lambda_{i,k}$, $\mu_A$ to each of them, one builds the Lagrangian function
\begin{equation}
{\pazocal L}(\boldsymbol{x}, \boldsymbol{y}, \boldsymbol{\psi}, \boldsymbol{\lambda}, \boldsymbol{\mu}) = \bra{\Psi_{\pazocal P}} {\pazocal B} \ket{\Psi_{\pazocal P}} + \sum_{i,k}\lambda_{i,k}g_{i,k} + \sum_A \mu_A h_A.
\label{eq:def:Lagrangian}
\end{equation}

In practice, one may want to fix one of the measurements, e.g. ${\pazocal M}_0^{(i)} = \sigma_x$ (equivalently, $x_i,0 = 1$ for all $i \in [n]$), since one can apply a local rotation to the measurements and the inverse rotation to the quantum state without changing the value of $\bra{\Psi_{\pazocal P}} {\pazocal B} \ket{\Psi_{\pazocal P}}$ \cite{AnnPhys}, thus reducing the number of variables in the optimization.

\cref{eq:def:Lagrangian} gives rise to a system of polynomial equations, where one of their solutions (the minimal among the real ones) will correspond to $\beta_k^{\pazocal P}$. B\'ezout's Theorem allows us to upper bound their number of solutions in the generic case \cite{Harris1992}. Recall that two algebraic curves of degrees $m$ and $n$ intersect at $m n$ points (which may be real, complex, or at infinity) and cannot meet at more than $mn$ points unless they have a common component, which generically will not be the case.

Finding the solutions to a system of polynomials consisting of $p$ equations and $p$ unknowns has been studied since long \cite{NoetherMathPhysKlasse1928}, yet it remains an area of intense research \cite{Sturmfels2002}. Since we have $p > 2$ equations, we have to look at generalizations of B\'ezout's theorem. Generalizations based on the homotopy method (see e.g. \cite{WrightMoC1985, GarciaSIAM1980, KojimaMP1983}) tell us that, generically, the number of solutions will be the product of the degrees of the $p$ polynomials.

 Hence, we need to count how many curves of which degree we have. On the one hand, we will have an equation for each variable involved in ${\pazocal L}$, which is $n+n+\sum_{A\in {\cal P}}2^{|A|}+n+|{\pazocal P}| \leq  3n + (n/k)2^k + n/k \leq n(2^k/k+4)$. On the other hand, the degree of every polynomial will be upper-bounded by the degree of ${\pazocal L}$, which is at most $n+2$. Hence, we obtain an upper bound to the number of solutions
\begin{equation}
(n+2)^{n(2^k/k+4)}.
\label{eq:BoundBezout}
\end{equation}

To find these solutions, one needs to calculate a Gr\"obner basis \cite{SturmfelsAMS2005, Sturmfels2002} in order to solve the system of polynomials arising from \eqref{eq:def:Lagrangian}. The computation of a Gr\"obner basis has, in the worst case, exponential complexity in the number of variables. In very pathological cases, this complexity can even become doubly-exponential \cite{FaugereF41999}. However, in practice, most computer algebra routines find them in times nowhere near these complexity bounds \cite{FaugereF41999, FaugereF52002}.

To conclude this section, let us briefly comment on the non-generic case. The most extreme and well-known example we can think of is that of a system of linear equations, where all the polynomial degrees are $1$, therefore one would expect generically one solution according to the above argument. Of course, systems of linear equations with non-trivial kernel arise in a variety of contexts. However, a random $\varepsilon$-perturbation of the system will bring it to the generic case. This shows how the underlying structure of polynomial systems is extremely rigid, giving rise in the worst-case to the above mentioned bounds. However, for the purposes of our work, one can always take an $\varepsilon$-perturbation of the coefficients $\alpha_{k_1,\ldots,k_p}^{(i_1,\ldots,i_p)}$ of the DIWED, if necessary. Furthermore, the aim of this section is to show that it is possible, in principle, to find exactly the $k$-producible bound of a DIWED of the form of \cref{eq:def:BellOp}.

\subsection{Two-body case}
\label{subsec:OptTwoBody}
In this section, we particularize the method shown in Section \ref{subsec:OptGeneric} to two-body Bell inequalities, and show that these give an exponential advantage in the number of solutions \eqref{eq:BoundBezout}.

Let us begin by writing the Bell operator ${\pazocal B}$ in this case:
\begin{equation}
 \label{eq:BellOp2}
 {\pazocal B} = \sum_{k \in [2]} \sum_{i \in [n]}  \alpha_k^{(i)} {\pazocal M}_k^{(i)} + \sum_{k, l\in [2]}  \sum_{i\neq j \in [n]} \alpha_{k,l}^{(i,j)}{\pazocal M}_k^{(i)} \otimes{\pazocal M}_l^{(j)}.
\end{equation}
When computing the quantity $\bra{\Psi_{\pazocal P}} {\pazocal B} \ket{\Psi_{\pazocal P}}$ it is now easy to see that each of the terms in ${\pazocal B}$ has support onto one or two of the states $\ket{\psi_A}$ that form $\ket{\Psi_{\pazocal P}}$. Therefore, expanding \cref{eq:BellOp2} yields in this case a polynomial of much lower degree, albeit with the same number of variables. In particular, the degree of $\bra{\Psi_{\pazocal P}} {\pazocal B} \ket{\Psi_{\pazocal P}}$ is now at most $2$ in the $\boldsymbol{x}$, $\boldsymbol{y}$ variables. Hence, all the equations stemming from \cref{eq:def:Lagrangian} have degree at most $4$, which yields an exponential improvement in the number of solutions, namely
\begin{equation}
4^{n(2^k/k+4)}.
\end{equation}

Therefore, if the Bell operator consists of two-body correlators, $\bra{\Psi_{\pazocal P}}{\pazocal B} \ket{\Psi_{\pazocal P}}$ can now be computed efficiently for a constant $k$ and large $n$, as it requires $O(2^{2k} n^2) = O(n^2)$ steps. In addition, note that in this case we can mildly relax the assumption that $k$ is a constant and still preserve the polynomial complexity of the algorithm: as long as $k \in O(\log n)$, the overall evaluation time will still be polynomial in $n$.

\subsection{The two-body symmetric case}
\label{subsec:OptTwoBodySymmetric}
In this section, we add the additional requirement that the Bell inequality $I$ should be permutationally invariant and built from one- and two-body correlators \cite{SciencePaper}. We shall see that the expectation value of the Bell operator can be computed in this case in a much more efficient way.
Motivated by the numerical results in \cref{sec:numerical}, in \cref{subsubsec:OptSymm} we also study the case in which all parties in the same region $A\in {\pazocal P}$ measure the same set of measurements, and when the state $\ket{\psi_{A}}$ is symmetric.

In the case of a permutationally invariant two-body Bell inequality, the coefficients $\alpha_k^{(i)}$ and $\alpha_{kl}^{(i,j)}$ in \cref{eq:BellOp2} do not depend on $i$ or $j$. Hence, the Bell operator can be expressed as
\begin{equation}
 \label{eq:BellOp2Sym}
 {\pazocal B} = \sum_{k \in [2]} \alpha_k \sum_{i \in [n]} {\pazocal M}_k^{(i)} + \sum_{k\leq l\in [2]} \alpha_{kl} \sum_{i\neq j \in [n]} {\pazocal M}_k^{(i)} \otimes{\pazocal M}_l^{(j)}.
\end{equation}
Its expectation value over $\ket{\Psi_{\pazocal P}}$ can be computed as
\begin{alignat}{2}
&\bra{\Psi_{\pazocal P}}{\pazocal B} \ket{\Psi_{\pazocal P}} \nonumber\\
 =& \sum_{{A}\in \pazocal{P}} \left( \sum_{k} \alpha_k \underbrace{\braket{\psi_{A}|{\pazocal{B}}_k^{A}|\psi_{A}}}_{\textmd{one-body terms}}+\sum_{k \leq l} \alpha_{kl} \underbrace{\braket{\psi_{A}|{\pazocal{B}}_{kl}^{A}|\psi_{A}}}_{\textmd{same region terms}} \right)\nonumber\\
+&\sum_{{A}\neq {A'}\in \pazocal{P}}\left(\sum_{k\leq l} \alpha_{kl}\underbrace{\braket{\psi_{A}|{\pazocal{B}^{A}_k}|\psi_{A}}\braket{\psi_{{A'}}|{\pazocal{B}^{{A'}}_l}|\psi_{{A'}}}}_{\textmd{crossed region terms}}\right),
\label{eq:Contraction}
\end{alignat}
where the terms of the Bell operator have been grouped in the following way:
\begin{equation}
 \label{eq:def:littleBs}
 {\pazocal B}_k^{A} := \sum_{i \in A}{\pazocal M}_k^{(i)},\quad {\pazocal B}_{kl}^{A} := \sum_{i \in A, i' \in A \setminus \{i\}} {\pazocal M}_k^{(i)}\otimes {\pazocal M}_l^{(i')}.
\end{equation}

\begin{figure}
  \includegraphics[scale=0.7]{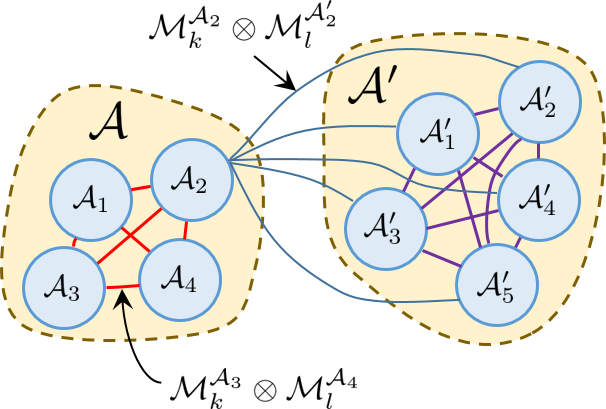}
  \caption{Interpretation of \cref{eq:Contraction}. We have represented two groups of parties, $\pazocal A$ and $\pazocal A'$, forming a partition ${\pazocal P}\in {\cal P}^k_{\mathrm{prod}}$ in the yellow shapes with discontinuous contour. Within each group there are several parties represented by the blue circles. The one-body terms of \cref{eq:Contraction} are not represented, as they correspond to each blue circle (party). The two-body terms are divided in two parts: Those that involve two parties belonging to the same group (red lines for $\pazocal A$, purple lines for $\pazocal A'$), and those that involve two parties that belong to different groups (Here, for clarity of the figure, we have only depicted the terms that connect ${\pazocal A}_2$ from ${\pazocal A}$ to any party in ${\pazocal A'}$, but of course there would be many more lines crossing different regions). Since the state $\ket{\Psi_{\pazocal P}}$ is product with respect to different regions, expectation values for the lines that connect different regions will factorize.
  }
  \label{fig:sameregionterms}
\end{figure}

In \cref{fig:sameregionterms} we graphically represent \cref{eq:Contraction}.

\subsubsection{Optimization over symmetric regions}
\label{subsubsec:OptSymm}

In this section, we are interested in finding an expression for the quantity $\bra{\Psi_{\pazocal P}}{\pazocal B} \ket{\Psi_{\pazocal P}}$ under the assumption that the Bell operator consists of two-body correlators and it is permutationally invariant on each region $A \in {\pazocal P}$. The last assumption is motivated by the numerical results that we shall present in Section \ref{subsec:variational}.

It is well-known that a permutationally-invariant operator $\pazocal B$ acting on a Hilbert space ${\pazocal H} = ({\mathbbm C}^{d})^{\otimes n}$ can be block-diagonalized into invariant subspaces that correspond to different spin components with their corresponding multiplicities, a result also known as Schur-Weyl duality. For a permutationally invariant operator of $n$ qudits, the size of these blocks is $O({n+d-1 \choose d-1})$ \cite{MoroderNJP2012, PhDTura, PhDHarrow, PhDChristandl, AudenaertNotes}.

For instance, consider an operator ${\pazocal B} \in {\cal B}(({\mathbbm C}^2)^{\otimes 2})$ acting on a two qubit space, which is permutationally invariant. Then, $\pazocal{B}$ is block-diagonalized in the basis spanned by the Dicke states and the anti-symmetric Bell state: $V^\dagger{\pazocal B}V = {\pazocal B}_{3\times 3} \oplus {\pazocal B}_{1 \times 1}$:
\begin{equation}
V^\dagger{\pazocal B} V = \left(
\begin{array}{ccc|c}
&&&\\
&{\pazocal B}_{3\times 3}&&0\\
&&&\\
\hline
&0&&{\pazocal B}_{1 \times 1}
\end{array}
\right), \ 
V = \left(
\begin{array}{ccc|c}
1&0&0&0\\
0&\frac{1}{\sqrt{2}}&0&\frac{1}{\sqrt{2}}\\
0&\frac{1}{\sqrt{2}}&0&\frac{-1}{\sqrt{2}}\\
0&0&1&0
\end{array}
\right).
\end{equation}

Let us recover the form of the Bell operator, from \cref{eq:Contraction}.
\begin{alignat}{2}
{\pazocal B} =& \sum_{A \in {\pazocal P}}\left(\sum_k \alpha_k {\pazocal B}_k^{A} + \sum_{k \leq l} \alpha_{kl}{\pazocal B}_{kl}^{A}\right)\nonumber\\
 +&\sum_{A\neq A'\in {\pazocal P}} \sum_{k \leq l} \alpha_{kl}{\pazocal B}_{k}^{A}\otimes {\pazocal B}_{l}^{A'}.
\label{eq:BellOp2Body}
\end{alignat}
Let $V_A$ be a matrix consisting of the Schur-Weyl basis vectors that block-diagonalize the $A$-th block arranged in columns. Then, by denoting $V = \bigotimes_{A\in {\pazocal P}} V_A$, we have that
\begin{alignat}{2}
V^\dagger{\pazocal B}V =& \sum_{A \in {\pazocal P}}\left(\sum_k \alpha_k V_A^\dagger{\pazocal B}_k^{A}V_A + \sum_{k \leq l} \alpha_{kl} V_A^\dagger{\pazocal B}_{kl}^{A}V_A\right)\nonumber\\
 +&\sum_{A\neq A' \in {\pazocal P}} \sum_{k \leq l} \alpha_{kl}(V_A^\dagger{\pazocal B}_{k}^{A}V_A)\otimes (V_{A'}^\dagger{\pazocal B}_{l}^{A'}V_{A'}).
\label{eq:BellOp2BodyBasisChange}
\end{alignat}
Note that $V_A V_A^\dagger = V_A^\dagger V_A = \mathbbm{1}_{{\pazocal H}_A}$. Therefore, any $k$-producible pure state $\ket{\Psi}$ is expressed as $V^\dagger {\ket \Psi}$ in the tensor product basis of the blocks.

It is of particular interest the case where $\ket{\psi_A}$ is fully supported on an invariant subspace of $V_A$ (e.g. the symmetric space of $A$). In this case, $\ket{\psi_A}$ is described by, at most, $|A|$ real parameters. Therefore, the number of local minima is upper bounded by
\begin{equation}
4^{3|{\pazocal P}| +n/k\cdot k} \leq 4^{n(1+3/k)}.
\end{equation}

If, in addition, all regions $|A|$ have the same cardinal $k$, and $\ket{\psi_A} = \ket{\psi_{A'}}$ for every $A,A' \in {\pazocal P}$, then the above bound is dramatically reduced to
\begin{equation}
4^{4+k}.
\label{eq:drama}
\end{equation}
This last case is of particular importance in the asymptotic analysis developed in \cref{sec:asymptotic}.

\section{The methods} \label{sec:methods}
In this section we describe the two main methods that allow us to construct the DIWEDs defined in \cref{eq:def:DIWED} from a particular two-body permutationally invariant Bell inequality of the form given in \cref{eq:def:PIBI}. In \cref{subsec:variational} we give a variational approach that allows to find a good upper bound $\beta_k^U$ to $\beta_k$. In \cref{subsec:sdp} we construct a certificate that provides a lower bound $\beta_{k}^L$ to $\beta_k$, thus complementing the bound given in \cref{subsec:variational}. Hence $\beta_{k}^L \leq \beta_k \leq \beta_k^U$.

\subsection{Variational upper bound} \label{subsec:variational}
To find an upper bound to the optimization given by \cref{eq:partitionfixed}, it is sufficient to find a $k$-producible vector $\ket{\Psi}$ and a set of measurements ${\pazocal M}_k^{(i)}$. Indeed, any expectation value provided by a $k$-producible state on some Bell operator upon performing any set of measurements leads, by construction, to an upper bound to $\beta_k$.

Our aim here is to describe in detail a method that gives a good guess to the optimal solution $\beta_k$, and we do so by constructing a $k$-producible pure state and finding a suitable set of measurements.
The steps of the algorithm are as follows:
\begin{enumerate}
 \item For every partition ${\pazocal P} \in {\mathcal P}_\mathrm{prod}^k$,
 \item pick a random $k$-producible pure state $\ket{\Psi}$ and random initial measurement settings, parametrized by $\boldsymbol \theta$.
 \item Use a see-saw iteration method (See e.g. \cite{PalPRA2010, WernerWolfQICReview2001}) to improve the choice of $\boldsymbol \theta$, while keeping $\ket{\Psi}$ fixed.

 The see-saw iteration technique consists in doing several sweeps across $[n]$ and optimizing the measurement parameters one site at a time. Thus, the see-saw iteration fixes the measurements ${\pazocal M}_k^{(i)}$ of all parties except for the $j$-th one. Then, one can write a cost function
 \begin{equation}
 E_j(\boldsymbol \theta_j) = \bra{\Psi}{\pazocal B}(\boldsymbol \theta_j; \boldsymbol \theta_0, \ldots, \hat{\boldsymbol \theta}_j, \ldots, \boldsymbol \theta_{n-1})\ket{\Psi}
 \label{eq:def:costfunctionseesaw}
 \end{equation}
 that only depends on $\boldsymbol \theta_j$ only, while the rest of them ($\boldsymbol \theta_i$, $i \neq j$) simply become parameters describing the fixed ${\pazocal M}_k^{(i)}$.
 
Finding $\boldsymbol \theta_j$ can be done using stochastic gradient descent methods. However, we note that, since \cref{eq:def:costfunctionseesaw} is linear in ${\pazocal M}_k^{(j)}$, it can be expressed as a semi-definite program whose variables are the positive-operator valued measure (POVM) elements of ${\pazocal M}_k^{(j)}$, thus obtaining the minimum of \cref{eq:def:costfunctionseesaw} in one iteration:

\begin{equation}
\begin{array}{lll}
\min_{\Pi_0^{(k)}}&& \mathrm{Tr}[\ket{\psi}\bra{\psi} {\pazocal B}(\boldsymbol{\theta})]\\
&\mathrm{s.t.}& 0 \preccurlyeq \Pi_0^{(k)} \preccurlyeq \mathbbm{1}\\
&&{\pazocal M_k^{(j)}} = \Pi_0^{(k)} - \Pi_1^{(k)}\\
&&\Pi_{0}^{(k)} + \Pi_1^{(k)} = \mathbbm{1}
\end{array}.
\label{eq:seesawsdp}
\end{equation}
 
We keep the value of the measurements stemming from \cref{eq:seesawsdp} as the updated value of $\boldsymbol{\theta}_j$, and we move to the next value of $j$. Note that the values of $\bra{\Psi}{\pazocal B}(\boldsymbol \theta)\ket{\Psi}$ form at each step a monotonically decreasing function, thus converging to a local minimum. We go to Step 4 once the sequence has converged within numerical accuracy.
 
 Alternatively, we note that the optimal choice of $\boldsymbol{\theta}$ minimizing $\bra{\Psi}{\pazocal B}(\boldsymbol \theta)\ket{\Psi}$ with $\ket{\psi}$ fixed, can be also found using stochastic gradient descent techniques. However, we discourage its use, as the number of free parameters grows with $n$ resulting in a worse performance than the see-saw approach, and a much higher sensitivity to the initial point $\boldsymbol{\theta}$.

 \item In this step we find a $k$-producible state that improves the value of $\bra{\Psi}{\pazocal B}(\boldsymbol \theta)\ket{\Psi}$, while keeping the measurement settings, parameterized by $\boldsymbol \theta$, fixed. Like in the previous step, a See-Saw iteration method is possible, by fixing the value of all $\ket{\psi_i}$, except for $i = j$, where $j$ enumerates the elements of $\pazocal P$. Note that in this case, \cref{eq:BellOp2Body} enables us to find $\ket{\psi_j}$ as the eigenvector of minimal eigenvalue of
 \begin{alignat}{2}
 \label{eq:def:Btilde}
  \tilde{\pazocal B_j} &= \sum_{k}\alpha_k{\pazocal B}_k^{A_j} + \sum_{k \leq l} \alpha_{kl} {\pazocal B}_{kl}^{A_j}\nonumber\\
  +& \sum_{k \leq l} \alpha_{kl}\sum_{A_i \neq A_j \in {\pazocal P}} \langle{\pazocal B}_k^{A_i}\rangle{\pazocal B}_l^{A_j} + {\pazocal B}_k^{A_j} \langle{\pazocal B}_l^{A_i}\rangle.\nonumber\\
\end{alignat}
 Like in step $3$, the value of $j$ is iterated back and forth until $\bra{\Psi} {\pazocal B}(\boldsymbol{\theta}) \ket{\Psi}$ converges within numerical accuracy.
 \item Repeat steps $3$ and $4$ until the value of $\bra{\Psi} {\pazocal B}(\boldsymbol{\theta}) \ket{\Psi}$ lies within numerical accuracy; i.e., the result from both steps $3$ and $4$ lie already within numerical accuracy.
 \item Define $\beta^U_k({\pazocal P}) = \bra{\Psi} {\pazocal B}(\boldsymbol{\theta}) \ket{\Psi}$. If there are partitions left, go to step $1$.
 
We note that it is not necessary to explore all ${\pazocal P}\in {\cal P}^k_{\mathrm{prod}}$, as one can define a partial order in ${\cal P}^k_{\mathrm{prod}}$ induced by inclusion: Formally, for every $\pazocal P$, $\pazocal Q$ $\in {\cal P}^k_{\mathrm{prod}}$, we say that
${\pazocal P} \preccurlyeq {\pazocal Q}$ if, and only if, for every $A \in {\pazocal P}$, there exists a $B \in {\pazocal Q}$ such that $A \subseteq B$. Then it is clear that $\beta_k^U({\pazocal P}) \geq \beta_k^U({\pazocal Q})$ if ${\pazocal P} \preccurlyeq {\pazocal Q}$.

Hence, since ${\cal P}^k_{\mathrm{prod}}$ is a poset, it is sufficient to pick $\pazocal P$ in step $1$ from its minimal elements.
 
 \item Output $\beta_k^{U} = \min_{\pazocal P} \beta^U_k({\pazocal P})$.
\end{enumerate}

As the end-result of the algorithm, we obtain a value $\beta_k^U \geq \beta_k$. Thanks to Jordan's lemma we do not need to increase the Hilbert space dimension of the states, nor consider measurements outside of the XZ plane. 
Recall, however, that the See-saw optimization method, although it allows to tackle multilateral optimization instances by splitting them into several unilateral approaches, therefore giving fast convergence rates, is a greedy algorithm which can be stuck at local minima. This depends highly on the landscape of the objective function and the number of local minima it has. In \cref{sec:kprod} we have discussed some of the worst-case bounds for the number of local minima, which gives an indication on the number of initial points necessary to ensure convergence with high probability.

\subsection{Lower bound certificate} \label{subsec:sdp}
Here we provide a complementary method to the one introduced in \cref{subsec:variational} to lower bound $\beta_k$ by means of a semidefinite program based on a relaxation of the quantum marginal problem.

Recall that the optimal value for $\beta_k$ is achieved at a certain partition $\pazocal P$, which we shall assume from here on to be fixed (we simply run the method for every minimal element of ${\cal P}^k_{\mathrm{prod}}$, as in \cref{subsec:variational} and keep the best value).
Let us recall that, given an arbitrary multipartite quantum state $\rho \in {\mathcal B}({\pazocal H})$, its expectation value on the Bell operator can be computed as (cf. \cref{eq:BellOp2Body})
\begin{alignat}{2}
\mathrm{Tr}[{\pazocal B}(\boldsymbol \theta) \rho]=&\nonumber\\
 \sum_{A_j \in {\pazocal P}}&\left(\sum_k \alpha_k \mathrm{Tr}[{\pazocal B}_k^{A_j}\rho_{A_j}] + \sum_{k \leq l} \alpha_{kl} \mathrm{Tr}[{\pazocal B}_{kl}^{A_j}\rho_{A_j}]\right)\nonumber\\
 +&\sum_{A_j\neq A_{j'}\in {\pazocal P}} \sum_{k \leq l} \alpha_{kl}\mathrm[{\pazocal B}_{k}^{A_j}\otimes {\pazocal B}_{l}^{A_{j'}}\rho_{A_j \cup A_{j'}}],
\label{eq:BellOp2BodySdP}
\end{alignat}
where $\rho_{A}$ is the reduced state of $\rho$ on the parties that constitute $A \subseteq [n]$.

Furthermore, $\rho$ is a $k$-separable state by assumption. Therefore, since $\pazocal P$ is fixed, $\rho$ must be separable across every cut $A_j | A_{j'}$. Therefore, we can find a lower bound to $\beta_k$ by considering the following problem
\begin{alignat}{3}
 \min \ &\mathrm{Tr}[{\pazocal B}(\boldsymbol \theta) \rho]\nonumber\\
 \mathrm{s. t. }\ & \rho_{A_j} \succeq 0,\ \rho_{A_j \cup A_{j'}} \succeq 0\nonumber\\
 &\mathrm{Tr}[\rho_{A_j}] = \mathrm{Tr}[\rho_{A_j \cup  A_{j'}}]=1\nonumber\\
 &\mathrm{Tr}_{A_j}[\rho_{A_j \cup A_{j'}}] = \rho_{A_j}\nonumber\\
 & \rho_{ A_j \cup A_{j'}} \mbox{ is separable across } A_j|A_{j'}.
 \label{eq:PreSdP}
\end{alignat}

The optimization presented in \cref{eq:PreSdP} is almost a semidefinite problem. Unfortunately, it is NP-hard to determine if a mixed quantum state is entangled or separable \cite{Gurvits2003}. However, the problem can be brought to a semidefinite programming form by allowing the feasible set to be an outer approximation to that of \cref{eq:PreSdP}.

An outer approximation to the set of separable states that behaves particularly well in semidefinite programming instances is the so-called positivity under partial transposition (PPT) \cite{PeresPRL1996}. Recall that a separable state $\rho$ can be expressed according to \cref{eq:def:separable}. Then, one can define its PPT version, which is the result of the map $({\mathbbm {1}}\otimes T)[\rho]=:\rho^\Gamma$, where $T$ is the transposition operator with respect to the computational basis. If $\rho$ is of the form of \cref{eq:def:separable}, then $\rho^\Gamma \succeq 0$.

Hence, we can now consider the relaxation of \cref{eq:PreSdP} in its semidefinite programming form:
\begin{alignat}{3}
 \min \ &\mathrm{Tr}[{\pazocal B}(\boldsymbol \theta) \rho]\nonumber\\
 \mathrm{s. t. }\ & \rho_{A_j} \succeq 0,\ \rho_{A_j \cup A_{j'}} \succeq 0\nonumber\\
 &\mathrm{Tr}[\rho_{A_j}] = \mathrm{Tr}[\rho_{A_j \cup  A_{j'}}]=1\nonumber\\
 &\mathrm{Tr}_{A_j}[\rho_{A_j \cup A_{j'}}] = \rho_{A_j}\nonumber\\
 & (\rho_{ A_j \cup A_{j'}})^\Gamma \succeq 0.
 \label{eq:SdP}
\end{alignat}

The relaxation arising from \cref{eq:SdP} can be tightened by considering stronger, yet numerically more expensive, approximations to the separable set \cite{HarrowCMP2017, DohertyPRA2004}. However, we shall see in \cref{sec:numerical} indications why it is not necessary to do so for inequalities of the form \cref{eq:def:PIBI} with $m=2$.

The convergence of \cref{eq:SdP} guarantees that the set of marginals with which we compute $\mathrm{Tr}[{\pazocal B}(\boldsymbol \theta) \rho]$ is actually a minimum. Although strong duality needs not hold in general for semidefinite programs, the cases where it does not hold are quite pathological \cite{ParriloBook2013}, weak duality (i.e., a feasible solution to the dual) would still give a lower bound to the minimum. \jordi{Check} Hence, for each choice of measurement settings, described by $\boldsymbol{\theta}$, we obtain a value $\beta_k^L(\boldsymbol{\theta}) \leq \beta_k^U(\boldsymbol{\theta})$. In particular, for the optimal choice of $\boldsymbol{\theta}$, we get a lower bound $\beta_k^L \leq \beta_k$.

By running also a see-saw optimization on $\boldsymbol{\theta}$, we find that the optimal value of $\beta_k^L(\boldsymbol{\theta})$ is obtained by the same value of $\boldsymbol \theta$ that yields the optimal value $\beta_k^U(\boldsymbol{\theta})$ whenever the initial value of $\boldsymbol \theta$ is close to the one yielding $\beta_k^U(\boldsymbol{\theta})$. Note that for values of $\boldsymbol \theta$ that are too far away from the optimal measurement parameters, it could happen that $\beta_k^L (\boldsymbol{\theta}') > \beta_k^U(\boldsymbol{\theta})$ (for instance, if $\boldsymbol{\theta}'$ corresponds to commuting observables, one can only produce the classical bound of the Bell inequality). However, one can discard all the measurement parameters yielding such a contradiction to also rapidly reduce the parameter space, in a similar spirit in which branch and bound methods operate: for instance, any $\boldsymbol{\theta'}$ giving a lower bound certificate strictly above any variational solution $\beta_k^{U}(\boldsymbol{\theta})$ automatically discards $\boldsymbol{\theta'}$ \cite{Baccari2018a}.

\section{Numerical results} \label{sec:numerical}
In this section, we present the numerical results we have calculated, obtaining DIWEDs for different PIBIs and number of parties. In \cref{subsec:unconstrained} we discuss the results obtained without any additional hypotheses, and we extract some properties that we conjecture are likely to be preserved in the many-body regime. In \cref{subsec:constrained} we perform the numerical analysis with these additional assumptions.

\subsection{Unconstrained optimization}\label{subsec:unconstrained}
We have performed numerical studies up to $n = 10$ parties using the see-saw method outlined in \cref{subsec:variational} with the following inequalities \cite{SciencePaper}:
\begin{equation}
 -2 {\pazocal S}_0 + \frac{1}{2}{\pazocal S}_{00} - {\pazocal S}_{01} +\frac{1}{2}{\pazocal S}_{11} \geq \beta_{k}
 \label{eq:ineq6}
\end{equation}
and
\begin{equation}
 (n \mbox{ mod } 2) (n-1) (n {\pazocal S}_0 + {\pazocal S}_1) + {n \choose 2} {\pazocal S}_{00} + n {\pazocal S}_{01} - {\pazocal S}_{11} \geq \beta_k.
 \label{eq:ineqdicke}
\end{equation}
The reason for these choices are the following:

Experiments have already been performed with \cref{eq:ineq6} \cite{SchmiedScience2016, EngelsenPRL2017}. Therefore, we can apply our method to draw conclusions about the entanglement depth that was present in such experiments. Concerning \cref{eq:ineqdicke}, it is an inequality that is tailored to the half-filled Dicke state \cite{AnnPhys}, therefore making it potentially relevant for experimental settings in which such states are prepared (e.g. via spin-changing collisions \cite{ExpLuecke2014}) and it is a generalization of CHSH \cite{CHSH} when $n=2$ is used.

To find $\beta_k$ for \cref{eq:ineq6} and \cref{eq:ineqdicke}, we have represented the quantum state in the computational basis (storing the whole state vector) and we have added no constraint among measurement settings, except setting ${\pazocal M}_0^{(i)} = \sigma_z$ for every $i \in [n]$. Therefore, the relevant measurement parameters are $\theta_i$ for each $i \in [n]$, thus parameterizing ${\pazocal M}_1^{(i)} = \cos \theta_i \sigma_z + \sin \theta_i \sigma_x$. This assumption incurs in no loss of generality, since the measurements of the $i$-th party can be simultaneously rotated to any measurements $\tilde{\pazocal M}_0^{(i)} = \cos(c_i) \sigma_z + \sin(c_i) \sigma_x$ and $\tilde{\pazocal M}_1^{(i)} = \cos(c_i + \theta_i) \sigma_z + \sin(c_i + \theta_i) \sigma_x$ by means of a local unitary rotation of angle $c_i$. This comes at the expenses of rotating the state with a single-qubit rotation of angle $-c_i$ applied to the $i$-th site, but this does not change the entanglement-depth properties of $\ket{\Psi_{\pazocal P}}$ \cite{AnnPhys}.

We have run the see-saw method with more than $10^4$ initial measurement settings, uniformly distributed in the $[0,2\pi]^n$ hypercube, and k-producible states, where the initial state at each region is chosen uniformly at random according to the Haar measure.

We observe the following behavior. For every partition $\pazocal {P}$ which is a maximal element of ${\cal P}^k_{\mathrm{prod}}$ with respect to the partial order $\preccurlyeq$, the optimal point is reached when the state at each region is permutationally invariant (in particular, it is a superposition of Dicke states), and the parties at each region perform the same measurement settings. The measurement settings among different regions, however are not always the same, but they  differ less if the size of these regions is similar.

Both the variational method and the complementary semidifinite programming certificate lead to the same results, within numerical accuracy; i.e., $\beta_k^U - \beta_k^L < 10^{-7}$. The numerical accuracy limitation is dominated by the accuracy of the SDP solver, which is of the order of $10^{-7}$. We have tried our methods with a variety of SDP solvers, including SeDuMi \cite{SeDuMi}, Mosek \cite{Mosek} and SDPT3 \cite{SDPT3}, with CVX \cite{CVX1, CVX2} and Yalmip \cite{YALMIP} as parsers, all leading to similar results.

It is worth noting that our results show a similar behavior to that observed in \cite{SciencePaper, AnnPhys}, where the optimal point is reached when all the parties in the same region  use the same $\theta_i$. The case treated in \cite{SciencePaper, AnnPhys} corresponds to the trivial partition ${\pazocal P} = \{[n]\}$, therefore our method recovers the original result.

\subsection{Optimization under additional hypotheses}\label{subsec:constrained}
Motivated by the behavior observed in \cref{subsec:unconstrained}, we have performed further numerical studies for $n, k \gg 10$, under the following assumptions:
\begin{itemize}
 \item The optimal measurement settings leading to $\beta_k$ are, modulo local unitary rotations, the same on each region ${A} \in {\pazocal P}$.
 \item The optimal state is of the form $\ket{\Psi_{\pazocal P}} = \ket{\psi_{A_1}}\otimes \cdots \otimes \ket{\psi_{A_{|{\pazocal P}|}}}$, where each  $\ket{\psi_A}$ is a superposition of symmetric states of the same spin length. In the case of \cref{eq:ineq6} and \cref{eq:ineqdicke}, this corresponds to states of the maximal spin length (Dicke states). These superpositions involve only real coefficients, therefore they can be described exactly with $|A|$ real parameters (including normalization).
\end{itemize}

In the case of \cref{eq:ineq6}, we observe that for $k$-producible bounds of large $k$, the state $\ket{\psi_A}$ can be well approximated (after suitable local unitary rotations \cite{AnnPhys}) by a superposition of Dicke states of the form
\begin{equation}
 \ket{\psi_A} \approx \sum_{j=0}^{|A|} c_j \ket{D_{|A|}^j},
 \label{eq:DickeSuperposition}
\end{equation}
where $c_j = e^{-(j-\mu_A)^2/4\sigma_A}/\sqrt[4]{2\pi \sigma_A}$ and $\ket{D_m^j}$ is the Dicke state of $m$ qubits with $j$ excitations.

\begin{figure}
\begin{center}
  \includegraphics[scale=0.45]{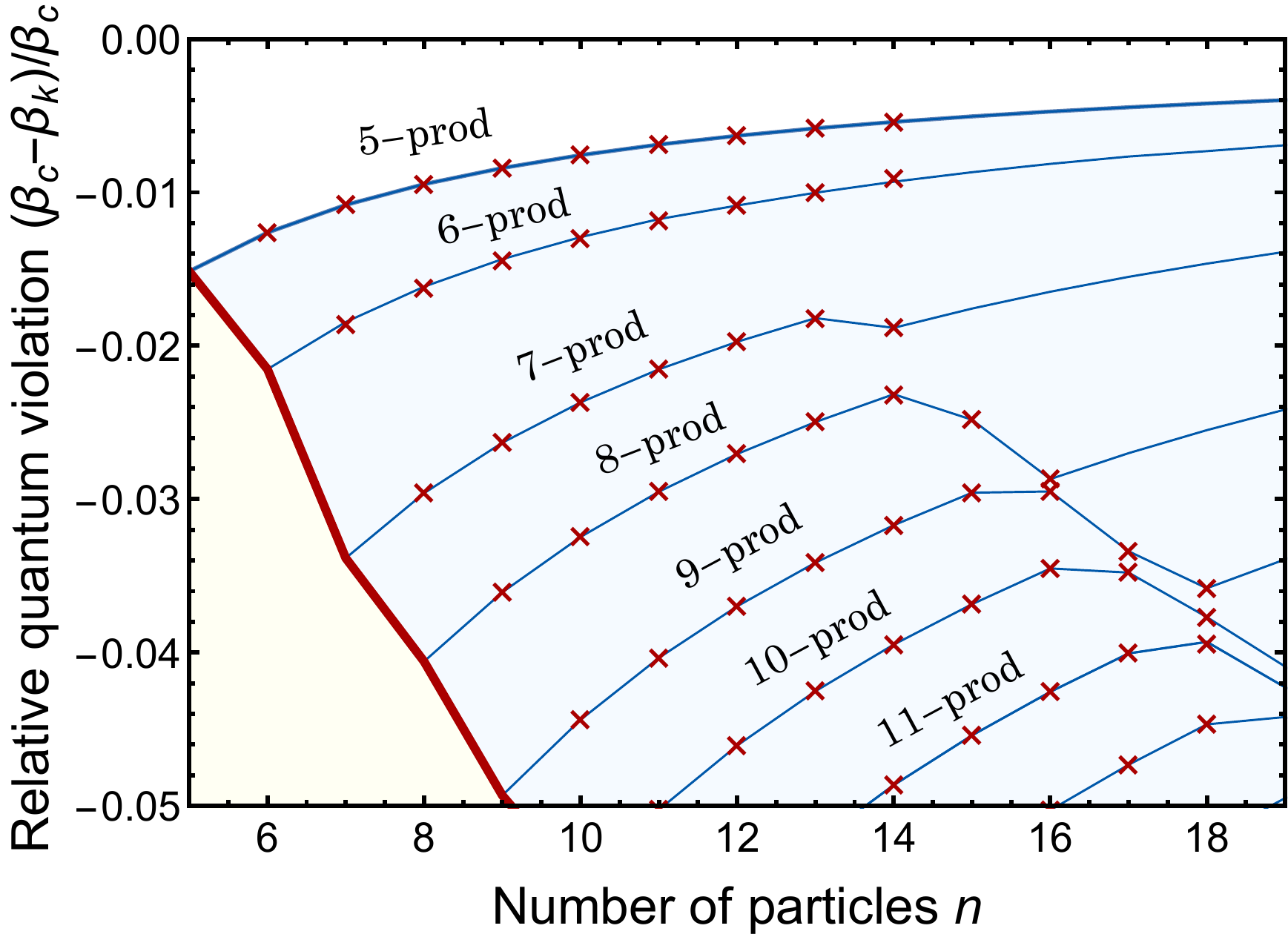}
  \caption{DIWED bounds for the 2-body PIBI presented in \cref{eq:ineq6} up to $20$ particles. Each line represents a $k$-producible bound. The wavy-like behavior of the bounds comes from the fact that \cref{eq:ineq6} has no quantum violation for less than $5$ parties \cite{SciencePaper}. Therefore, the optimal partition $\pazocal P$ for every $(n, k)$ tries to avoid forming groups of $4$ parties or less. The lines correspond to the variational solution (connected for better visibility) and the crosses to the lower bound certificate obtained via a semidefinite program. We see that for practical effects, both values always coincide.}
  \label{fig:comparissonMethodsIneq6}
  \end{center}
\end{figure}

We believe this behavior arises due to the following reasons: \cref{eq:ineq6} is maximally violated by this kind of states, and the maximal quantum violation, normalized to the classical bound, is monotonically increasing in magnitude, tending to a constant value. Therefore, the optimization to find $\beta_k$ favors this family of states on each region, which become better as the size of the region increases. In addition, \cref{eq:ineq6} has no quantum violation for less than $5$ parties. This gives rise to the wave-like behavior for small values of $n$ in \cref{fig:comparissonMethodsIneq6} (see also \cite{Aloy2018}). Furthermore, we see that due to this reason, larger regions give the optimal value of $\beta_k$, for which we conjecture that the most balanced partition (every $A \in {\pazocal P}$ has $k$ elements, except perhaps for one if $n$ is not a multiple of $k$) is the one that produces $\beta_k$. We analyse the asymptotic behavior of the DIWED arising from \cref{eq:ineq6} under this assumption in \cref{sec:asymptotic}.

\begin{figure}[h!]
\begin{center}
  \includegraphics[scale=0.37]{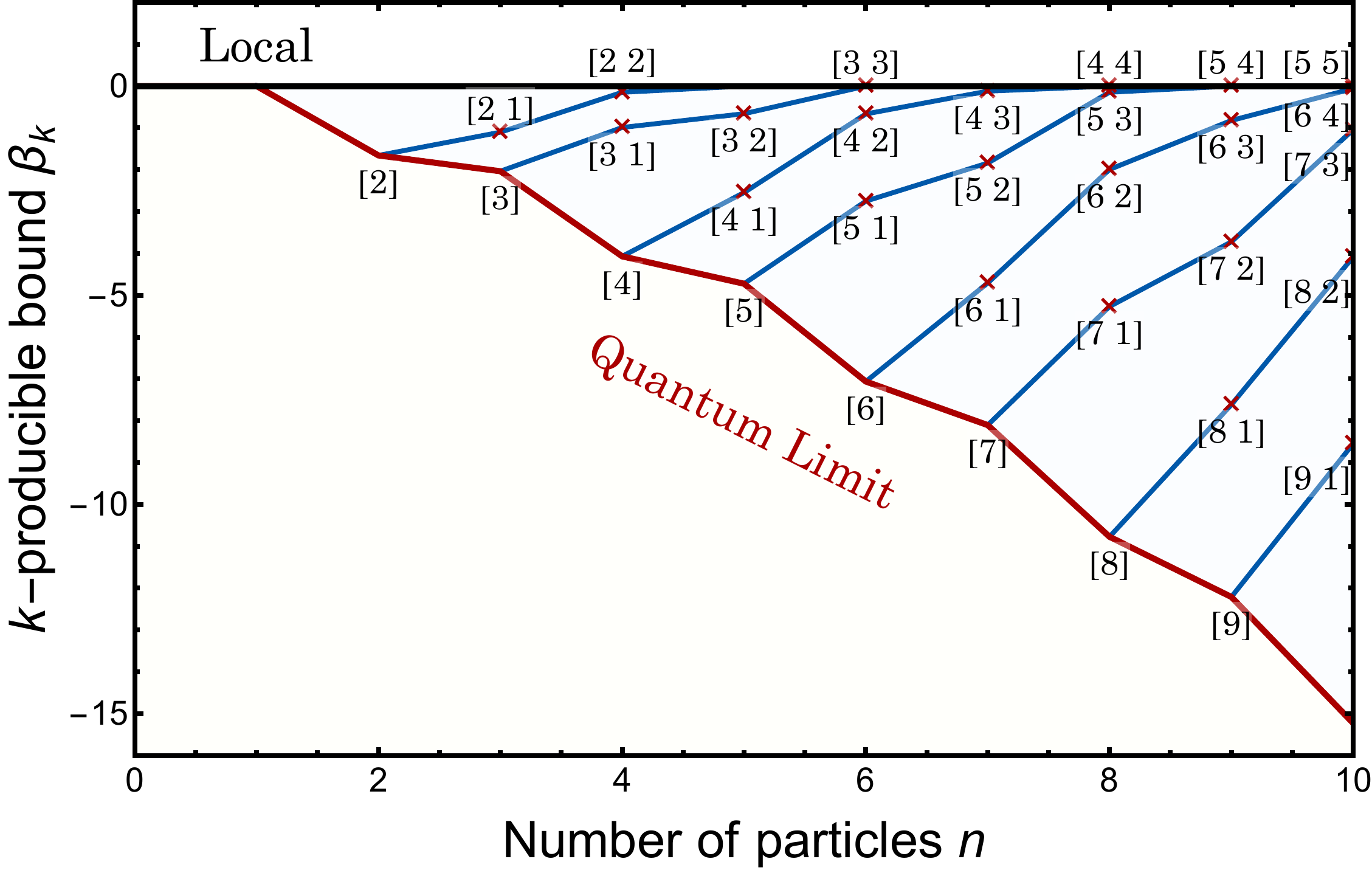}
  \caption{Entanglement depth bounds for Inequality (\ref{eq:ineqdicke}) up to 10 particles. Each line represents a $k$-producible bound and each cross the corresponding lower-bound certificate. We also label the partition yielding $\beta_k$ next to the corresponding point in the plot. In some cases, different partitions yield the same bound, for instance $\{7,2\}$ and $\{7,1,1\}$. We only label one of them for clarity. We observe that both methods converge up to numerical precision.}
  \label{fig:chsh}
  \end{center}
\end{figure}

\begin{figure}[h!]
\begin{center}
  \includegraphics[scale=0.3]{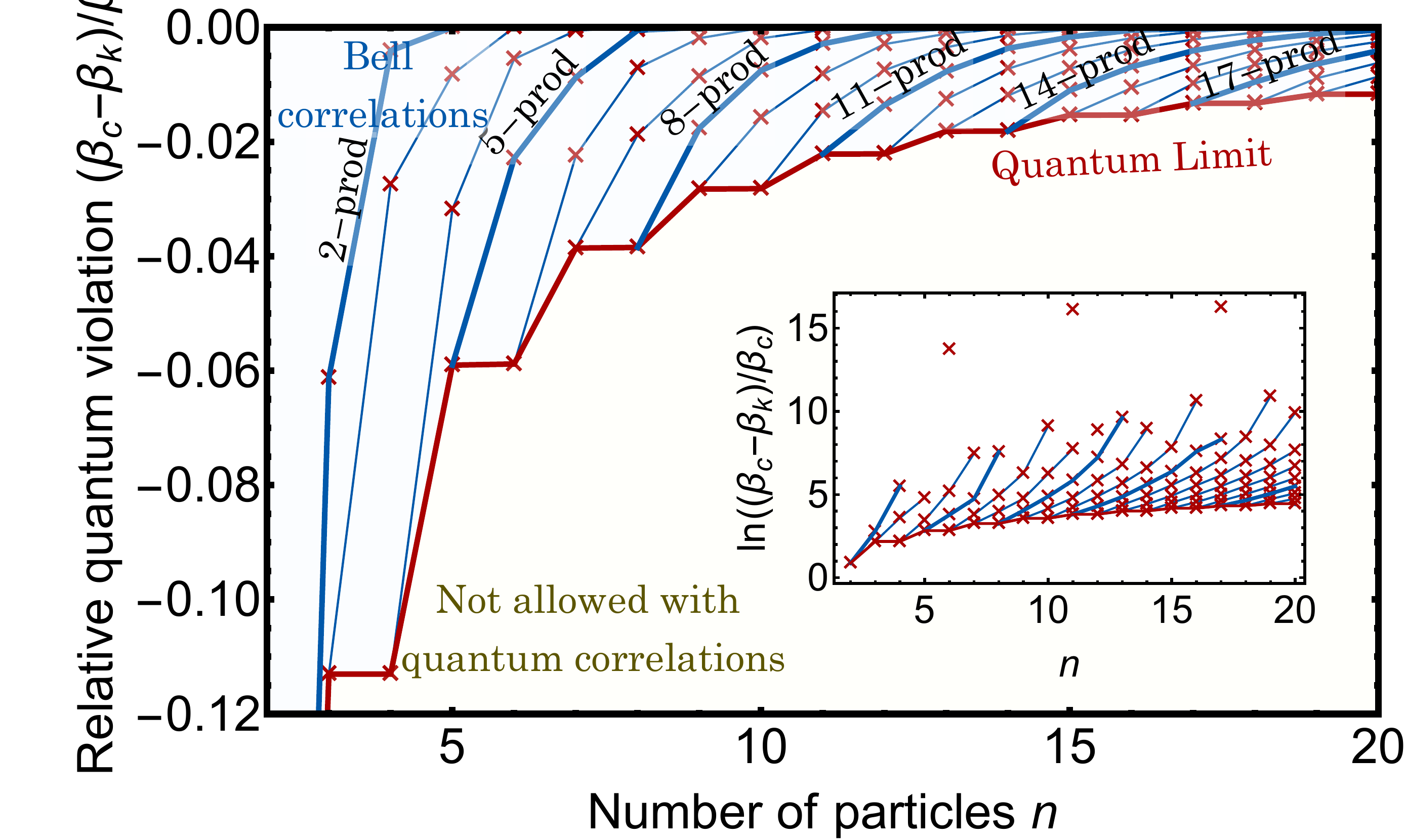}
  \caption{Entanglement depth bounds for Inequality (\ref{eq:ineqdicke}) up to 20 particles relative to the classical bound. The inset shows how the precision required to certify entanglement depth increases with the number of particles, as the relative quantum limit approaches zero with the system size. In some cases, the variational method would give zero, whereas the lower bound certificate would yield a value close to zero with negative sign, due to numerical precision. This explains the three isolated points in the inset.}
  \label{fig:chshnormalized}
  \end{center}
\end{figure}

In the case of \cref{eq:ineqdicke}, its behavior is shown in \cref{fig:chsh}, and normalized to the classical bound in \cref{fig:chshnormalized}. In this case, we see that the optimal partition is more difficult to predict, what we attribute to the fact that the relative quantum violation of \cref{eq:ineqdicke} decreases with the system size \cite{SciencePaper, AnnPhys}, although the optimal state tends asymptotically to the half-filled Dicke state $\ket{D_n^{\lceil n/2 \rfloor}}$.


\section{Asymptotic analysis} \label{sec:asymptotic}

In this section we perform an asymptotic analysis of the DIWED arising from \cref{eq:ineq6}, in the limit of large $n$ with the following assumptions, motivated by our findings in \cref{sec:numerical}:
\begin{itemize}
 \item The optimal partition ${\pazocal P} \in {\cal P}^k_{\mathrm{prod}}$ is given by $\lfloor k/n\rfloor$ groups of $k$ parties, plus a smaller partition of $n \mod k$ elements, if $k$ does not divide $n$.
 \item The optimal measurement settings on each group of parties are the same for each region (up to local unitaries).
 \item The measurement settings on different regions converge to the same (up to local unitaries) as the system becomes bigger.
\end{itemize}
Let us remark that the DIWED arising from \cref{eq:ineq6} does not detect entanglement depths for $k<5$ due to the nature of \cref{eq:ineq6}, which has no quantum violation for less than $5$ parties. Moreover, the contribution of the smaller group fades away as $n$ grows, so we will consider $k$ to be a multiple of $n$ for large values of $n$. Therefore, we define $m = |{\pazocal P}| = n/k$. Note that there is no loss of generality in ignoring the asymptotic analysis of the cases where $m$ does not divide $n$; this has also been observed in other works (see e.g. \cite{FroewisNatComms2017}).

Since we assume all parties to perform the same set of measurements, we can write $\boldsymbol{\theta} \equiv (\varphi, \theta)$, which parametrize the measurement operators as ${\pazocal M}_0^{(i)} = \cos \varphi \sigma_z^{(i)} +\sin \varphi \sigma_x^{(i)}$ and ${\pazocal M}_1^{(i)} = \cos \theta \sigma_z^{(i)} +\sin \theta \sigma_x^{(i)}$. There will be a value of $\theta - \varphi$ in which the state at each region can be parametrized as \cref{eq:DickeSuperposition}, with $\sigma_A =: \sigma$ and $\mu_A =: \mu$ being independent of $A\in {\pazocal P}$.

Let us begin by noting the following decomposition of the expectation value of a two-body Bell operator of the form of \cref{eq:BellOp2} on a $k$-producible state $\ket{\Psi_{\pazocal P}}$:
\begin{alignat}{2}
 \langle {\pazocal B} \rangle = &\sum_{A \in {\pazocal P}} \langle {\pazocal B}^A \rangle \nonumber\\
 + \sum_{A \neq A' \in {\pazocal P}} &\frac{\gamma}{2} \langle {\pazocal S}_0^A \rangle \langle {\pazocal S}_0^{A'} \rangle + \delta \langle {\pazocal S}_0^A \rangle \langle {\pazocal S}_1^{A'} \rangle + \frac{\varepsilon}{2}\langle {\pazocal S}_1^A \rangle \langle {\pazocal S}_1^{A'} \rangle,
\end{alignat}
where ${\pazocal S}_k^A$ and ${\pazocal B}^A$ are the respective restrictions of ${\pazocal S}_k$ and $\pazocal B$ on $A$.

In particular, for the particular Bell inequality \cref{eq:ineq6} we have
\begin{alignat}{2}
 \langle {\pazocal B} \rangle = &\sum_{A \in {\pazocal P}} \langle {\pazocal B}^A \rangle \nonumber\\
 +& \frac{1}{2}\sum_{A \neq A' \in {\pazocal P}}  (\langle {\pazocal S}_0^A \rangle - \langle {\pazocal S}_1^A \rangle)(\langle {\pazocal S}_0^{A'} \rangle - \langle {\pazocal S}_1^{A'} \rangle).
 \label{eq:BellOpIneq6Schneidet}
\end{alignat}

Now we need to recall two results from \cite{AnnPhys}:
\begin{itemize}
 \item The quantum limit $\beta_n$ of \cref{eq:ineq6} for a system of $n$ parties behaves asymptotically as
 \begin{equation}
  \beta_n = -\frac{5}{2}n + \frac{\sqrt{3}}{2} n^{1/2} - \frac{3}{2} + o(1).
  \label{eq:quantumlimit}
 \end{equation}
 The derivation of \cref{eq:quantumlimit} uses a refinement from the approximation made in \cite{AnnPhys}. Here we use that the $\sigma$ parameter in \cref{eq:DickeSuperposition} can be taken as $\sigma = \sqrt{n/48}$ for large $n$ and not just $\Theta(\sqrt{n})$. The details can be found in \cref{app:asymptotic}.
 \item The one-body reduced density matrix corresponding to the optimal state can be approximated as
 \begin{equation}
 \rho_1 = \frac{n}{(n-1)} \ket{+}\bra{+} + \left(
  \begin{array}{cccc}
   \frac{-1-2c}{2(n-1)}&0\\0&\frac{2c-1}{2(n-1)}
  \end{array}\right) + o(1),
  \label{eq:onebodyrho}
 \end{equation}
 where $c = \mu - n/2 = 1/(4\cos \theta)$ and $\ket{+}=(\ket{0} + \ket{1})/\sqrt{2}$.
\end{itemize}

It is now clear how to approximate the $k$-producible bound for large $n$ and $k$, by combining \cref{eq:BellOpIneq6Schneidet} with \cref{eq:quantumlimit} and \cref{eq:onebodyrho}. For the sake of clarity, we use the following measurement parameters: $\varphi = \pi - \theta$, $\theta = 5 \pi/6$.

\begin{figure}
\begin{center}
  \includegraphics[scale=0.5]{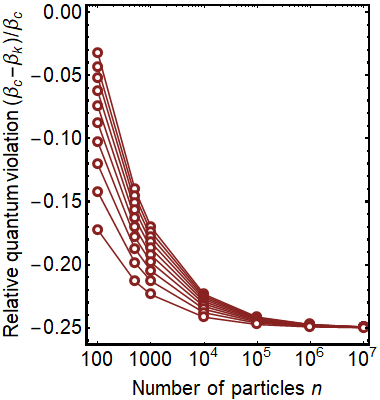}
  \caption{Asymptotic $k$-producibility bounds for the DIWED in \cref{eq:ineq6} for $k=n/m$, where $m$ is an integer ranging from $1$ (lowest line, coinciding with the maximal quantum violation of the Bell inequality) to $10$. In case of not exact divisibility, the value of $n$ that is a multiple of $m$ and is closest to a power of $10$ has been chosen. For this asymptotic computation, we have taken a state of the form \cref{eq:DickeSuperposition} and we have numerically optimized $\mu_A$ and $\sigma_A$. The optimal value converges to all $\mu_A$ and $\sigma_A$ equal, since all elements in $\pazocal P$ have the same cardinality. We also observe that taking $m = O(1)$ does not allow in the large $n$ limit for an experimentally robust result, and a different scaling is therefore needed, as shown in \cref{fig:asymptotictrend2}.}
  \label{fig:asymptotictrend1}
  \end{center}
\end{figure}

\begin{figure}
{\includegraphics[scale=0.5]{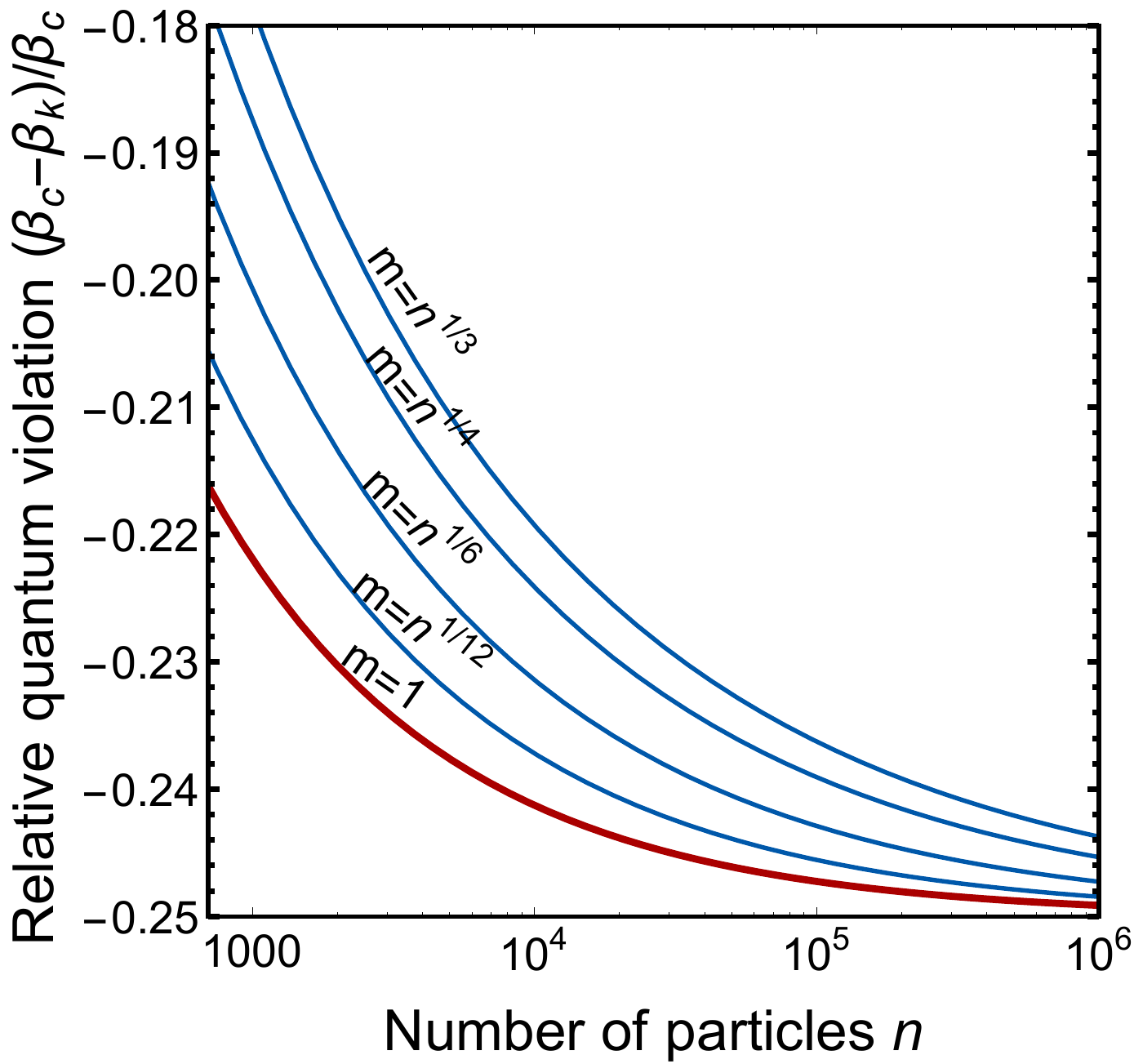}}
  \caption{Asymptotic bounds of the DIWED stemming from \cref{eq:ineq6} for different values of $k=n/m$, according to \cref{eq:asymptresult}. Although all the curves converge to the maximal quantum violation in the thermodynamic limit, these curves would allow for entanglement depth detection in experimentally realistic parameter regimes \cite{SchmiedScience2016, EngelsenPRL2017}.}
  \label{fig:asymptotictrend2}
\end{figure}

On the one hand, the expectation value $\langle {\pazocal B}^A \rangle$ on a region of size $k$ will asymptotically converge to
\begin{equation}
 \langle {\pazocal B}^A \rangle = -\frac{5}{2}k + \frac{\sqrt{3}}{2} k^{1/2} -\frac{3}{2} + o(1).
 \label{eq:approxregion}
\end{equation}
On the other hand, the expectation values of $\langle {\pazocal S}_0^A \rangle$ and $\langle {\pazocal S}_1^A \rangle$ will asymptotically tend to
\begin{equation}
 \langle {\pazocal S}_0^A \rangle = k \left(2c\cos(\theta)/(k-1) + \sin \theta k/(k-1)\right)
\end{equation}
and
\begin{equation}
 \langle {\pazocal S}_1^A \rangle = k \left(-2c\cos(\theta)/(k-1) + \sin \theta k/(k-1)\right).
\end{equation}
Therefore, with the optimal asymptotic value of $\theta = 5\pi/6$ we obtain
\begin{equation}
 \langle {\pazocal S}_0^A \rangle - \langle {\pazocal S}_1^A \rangle = \frac{k}{k-1}.
\end{equation}
Hence, the asymptotic limit for a partition of $m=n/k$ elements reads
\begin{equation}
 \beta_k \approx m\left(-\frac{5}{2}k+\frac{\sqrt{3}}{2}k^{1/2}-\frac{3}{2}\right) + {m \choose 2}\left(\frac{k}{k-1}\right)^2.
\end{equation}
In terms of its relative quantum violation $\tilde{\beta_k}:= (\beta_c - \beta_k)/\beta_c$, we have that
\begin{equation}
 \tilde{\beta_k} = -\frac{1}{4} + \frac{\sqrt{3m}}{4} n^{-1/2} + \frac{m^2-4m}{4} n^{-1} + O(n^{-3/2}).
 \label{eq:asymptresult}
\end{equation}
In \cref{fig:asymptotictrend1} and \cref{fig:asymptotictrend2} we have shown the bounds obtained for large $n$ and different $k$-producibility bounds, where $k=n/m$ and $m\in [10]$ (\cref{fig:asymptotictrend1}) or $m=n^{1/x}$ for several integer values of $x$ (\cref{fig:asymptotictrend2}). We see that as $n$ grows the bounds tend to the relative maximal quantum violation $-1/4$ as $1/\sqrt{n}$, and the approximation holds as long as $m = O(\sqrt[3]{n})$. Note that for much larger values of $m$ (e.g., $m=O(n)$), this corresponds to a small value for $k$, for which the approximation in \cref{eq:approxregion} could break down.

\section{Experimental realizations} \label{sec:experimental}
Although the DIWEDs introduced in this work allow to certify certain entanglement depths by performing a Bell test on spatially separated parties, a Bell test in the many-body regime is much more demanding. The full potential of our DIWEDs can be unlocked by means of Bell correlation witnesses, which only require trusted collective measurements \cite{SchmiedScience2016}, thus certifying the depth of entanglement in the state without an explicit characterization of the latter. The goal of this section is to rewrite our DIWEDs in terms of trusted collective measurements, which can be already performed in current experiments \cite{SchmiedScience2016, EngelsenPRL2017}.

When all parties perform the same measurements, it is clear that ${\pazocal S}_k$ can be directly mapped to a collective spin measurement along the direction of ${\pazocal M}_k$, and it is also easy to see that the expectation value of ${\pazocal S}_{kk}$ can be obtained by measuring the second moments of such collective spin measurement along the direction of ${\pazocal M}_k$ \cite{AnnPhys}. The reason for that is simply that one can rewrite ${\pazocal S}_{kk} = ({\pazocal S}_k)^2-n$. The ${\pazocal S}_{01}$ term is, however, more problematic, since one would apparently need to individually address the different parties in the system in order to measure it. One possibility is to estimate the second moments of the collective spin component along the directions given by $({\pazocal M}_0 \pm {\pazocal M}_1)/\sqrt{2}$ \cite{AnnPhys}, but this approach could be somewhat unsatisfactory, as evaluating the expectation value of the Bell inequality then requires measuring along two extra directions.

Let us briefly recall the approach in \cite{SchmiedScience2016} in order to show how one would measure a DIWED in the many-body regime (specifically, in a Bose-Einstein condensate (BEC)) using trusted collective measurements. This will allow us to compare the performance of our DIWEDs to other existing entanglement and nonlocality criteria, as we show in \cref{fig:comparison} and in \cref{sec:otherent}. Note that the Bell inequality that was actually used in \cite{SchmiedScience2016} belongs to the same equivalence class as \cref{eq:ineq6}, as one can take all its coefficients to be non-negative by renaming measurement outcomes appropriately \cite{AnnPhys}. Here we construct the entanglement depth witness for \cref{eq:ineq6} to be consistent with the rest of the paper. %

Using the same hypotheses as in \cite{SchmiedScience2016}; namely, that the system is characterized by a quantum mechanical description, that the experimental calibration of the measuremetns is trusted and that the particles do not communicate nor interact through channels unaccounted for, one can derive an entanglement depth witness that bypasses the problems due to the ${\pazocal S}_{01}$ term. The steps that we now follow are essentially identical to the ones in \cite{SchmiedScience2016}, substituting the classical bound $\beta_c$ by the $k$-producible bound $\beta_k$, and changing one of the signs of the measurement directions, due to the inequalities considered here and in \cite{SchmiedScience2016}  being two different representatives of the same equivalence class. Nevertheless, in the interest of the current section being self-contained, we reproduce them here for completeness.

First, one associates to the $i$-th observer a spin-$1/2$ particle and considers that the measurements are spin projections along a direction $\boldsymbol{d}$ in the Bloch sphere; i.e., ${\pazocal M}_{\boldsymbol{d}}^{(i)} = 2 \hat{s}^{(i)}\cdot \boldsymbol{d} \equiv \boldsymbol{\sigma}^{(i)} \cdot \boldsymbol{d}$, where $\boldsymbol{\sigma}^{(i)} := \{\sigma_x^{(i)},\sigma_y^{(i)},\sigma_z^{(i)}\}$ formally represents a vector formed by the Pauli matrices, and $\hat{s}^{(i)}$ is the individual spin operator acting on the $i$-th site. It is worth noting that the spin-$1/2$ description corresponds to the lowest energy levels of the atoms, and that higher energy levels as well as further degrees of freedom such as atomic motion are neglected in this model of description.

Second, one defines the total spin observable as the sum of the individual spins along a given direction: $\hat{S}_{\boldsymbol{d}} = \boldsymbol{d} \cdot \sum_{i} \hat{s}^{(i)}$. Consider now two unit vectors $\boldsymbol{a}$ and $\boldsymbol{n}$ and define a new vector $\boldsymbol{m} = 2(\boldsymbol{a} \cdot \boldsymbol{n}) \boldsymbol{a} - \boldsymbol{n}$. We note that $\boldsymbol{m}$ is also a unit vector and it is the result of applying a symmetry transformation to $\boldsymbol{n}$ with respect to the symmetry axis defined by $\boldsymbol{a}$.
By setting the measurement directions to be ${\pazocal M}_0^{(i)} = {\pazocal M}_{\boldsymbol{n}}^{(i)}$ and ${\pazocal M}_1^{(i)} = {\pazocal M}_{\boldsymbol{m}}^{(i)}$, the interpretation of $\boldsymbol{a}$ is clear, as it is the bisector of the angle formed between the two measurement directions.

In terms of the total spin component, we note the identities $\langle \hat{S}_{\boldsymbol{n}} \rangle = \langle {\pazocal S}_0 \rangle / 2$, and $16 (\boldsymbol{a} \cdot \boldsymbol{n})^2\langle \hat{S}_{\boldsymbol{a}}^2 \rangle = \langle {\pazocal S}_{00} \rangle + 2\langle {\pazocal S}_{01} \rangle + \langle {\pazocal S}_{11} \rangle + 4n (\boldsymbol{a} \cdot \boldsymbol{n})^2$.

Now, we would like to obtain a term of the form $\langle {\pazocal S}_{00} \rangle - 2\langle {\pazocal S}_{01}\rangle + \langle {\pazocal S}_{11} \rangle$, in order to map the two-body terms of \cref{eq:ineq6} to the Bell correlation witness we are constructing. This corresponds to the symmetry ${\pazocal M}_0^{(i)} \leftrightarrow -{\pazocal M}_{0}^{(i)}$ (renaming the outcomes of the $0$-th observable of every party). Thus, let us define $\boldsymbol{d} = -{\boldsymbol n}$. Then $\boldsymbol{m} = -2 (\boldsymbol{a}\cdot \boldsymbol{d})\boldsymbol{a}+\boldsymbol{d}$ and in a similar fashion we arrive at the new identity, $16(\boldsymbol{a}\cdot \boldsymbol{d})^2 \langle \hat{S}_{\boldsymbol{a}}^2\rangle = \langle {\pazocal S}_{00} \rangle - 2\langle {\pazocal S}_{01} \rangle + \langle {\pazocal S}_{11} \rangle + 4n (\boldsymbol{a} \cdot \boldsymbol{d})^2$

Hence, by normalizing to the classical bound, we can now turn our DIWED \cref{eq:ineq6} into the following witness, which only consists of first and second moments of (trusted) collective spin measurements:
\begin{equation}
\hat W = -2 \hat{S}_{\boldsymbol d}/n  + 4 (\boldsymbol{a}\cdot \boldsymbol{d})^2 \hat S_{\boldsymbol a}^2/n - (\boldsymbol{a}\cdot\boldsymbol{d})^2 - \frac{\beta_k}{2n}{\mathbbm{1}},
\label{exp:diwed}
\end{equation}
which satisfies the inequality $\langle \hat W \rangle \geq 0$ on $k$-producible states. Thus, an expectation value $\langle \hat W \rangle < 0$ certifies an entanglement depth of at least $k+1$ particles being genuinely multipartite entangled (see \cref{fig:comparison}).

\section{Comparison to other entanglement criteria} \label{sec:otherent}
\begin{figure}[h!]
\begin{center}
  \includegraphics[scale=0.6]{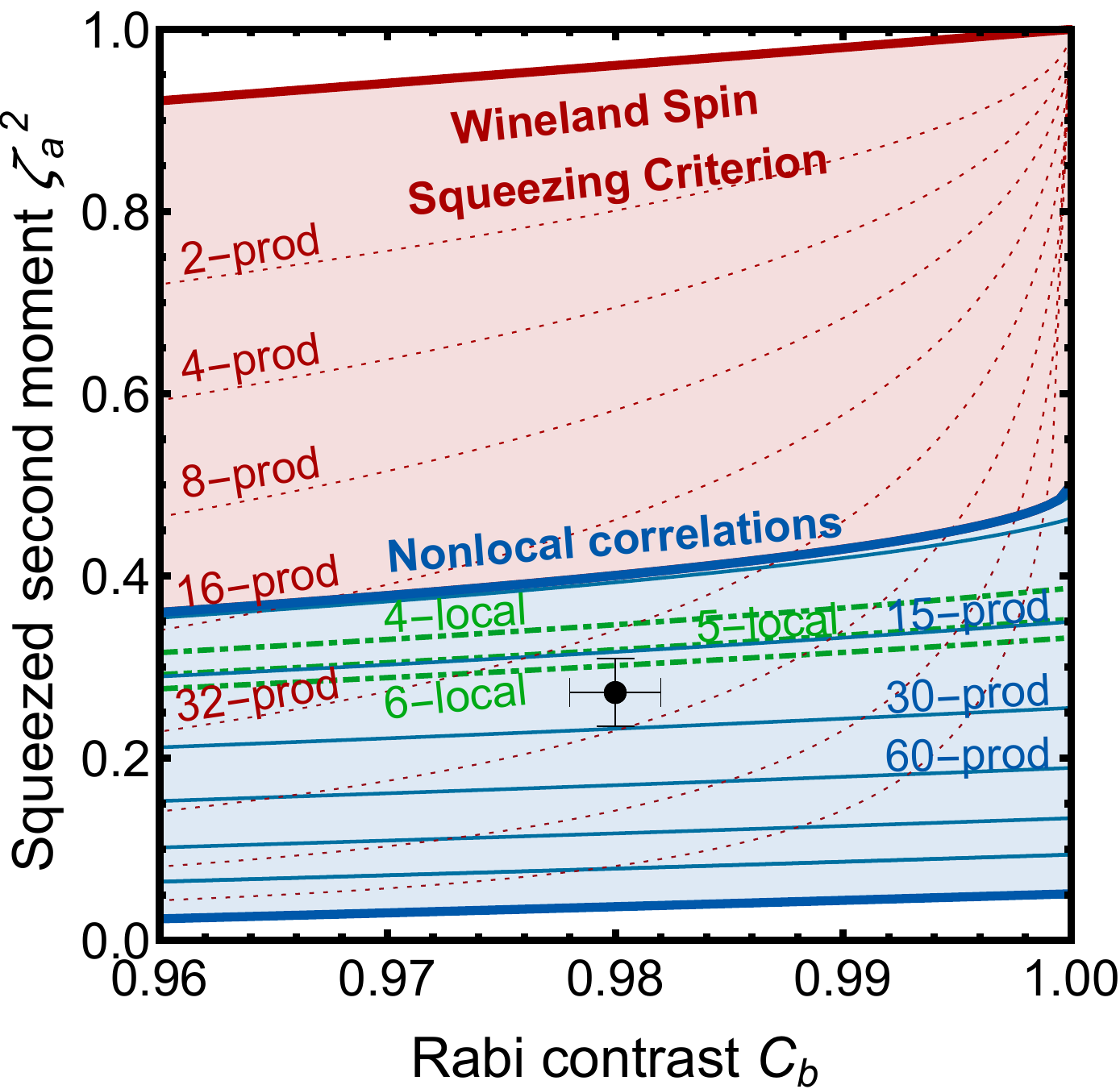}
  \caption{Witnesses of entanglement depth derived from trusted collective measurements. The blue lines follow \cref{eq:blueline} for different values of $k$. In green we have plotted the nonlocality depth bounds for \ref{eq:ineq6} for the values of $k=4,5,6$. Finally, in red, the not device-independent witnesses of entanglement depth derived from the Wineland spin squeezing criterion.}
  \label{fig:comparison}
  \end{center}
\end{figure}

In order to compare with other entanglement criteria, let us recall the definition of the scaled collective spin, also known as spin contrast, in the direction $\boldsymbol{d}$: $C_{\boldsymbol{d}}:= \langle 2 \hat S_{\boldsymbol d}  / \hat{n}\rangle$, where $\hat{n}$ is the particle number operator. Note that one has to take into consideration that the number of particles may vary among different experimental realizations \cite{SchmiedScience2016}. However, $[\hat{S}_{\boldsymbol{d}}, \hat{n}] = 0$, thus allowing to rewrite $C_{\boldsymbol{d}}= \langle 2 \hat S_{\boldsymbol d} \rangle / n$. Let us also recall the definition of the scaled second moment collective spin in the direction $\boldsymbol{a}$, $\zeta_{\boldsymbol{a}}^2 := \langle 4 \hat{S}_{\boldsymbol{a}}^2  / \hat{n} \rangle  =\langle 4 \hat{S}_{\boldsymbol{a}}^2\rangle/n$ for the same reason. Therefore, we have that the expectation value of $\hat W$ defined in \cref{exp:diwed} can be recast as
\begin{equation}
\langle \hat{W} \rangle = - C_{\boldsymbol{d}} + (\boldsymbol{a}\cdot \boldsymbol{d})^2 \zeta_{\boldsymbol a}^2 - (\boldsymbol{a}\cdot\boldsymbol{d})^2 - \frac{\beta_k}{2n},
\label{exp:diwed:expectationvalue}
\end{equation}
In this form, an algebraic manipulation \cite{SchmiedScience2016} allows us to compare the performance of our DIWED with other entanglement criteria such as Wineland's \cite{WinelandPRA1994, SorensenPRL2001} (see \cref{fig:comparison}). In order to do so, let us decompose the direction $\boldsymbol{d}$ into three orthonormal vectors $\boldsymbol{a}, \boldsymbol{b}, \boldsymbol{c}$ such that $\boldsymbol{a}$ is the squeezed axis (hence, we can assume $C_{\boldsymbol{a}} \approx 0$ and we only need to sweep across the $\boldsymbol{b}$ direction). Therefore, the inequality $\langle \hat W \rangle \geq 0$ becomes (cf. \cref{exp:diwed:expectationvalue})
\begin{equation}
\zeta_{\boldsymbol{a}}^2 \geq \frac{1}{2}\left(1 + \frac{\beta_k}{2n} - \sqrt{\left(1-\frac{\beta_k}{2n}\right)^2- C_{\boldsymbol{b}}^2}\right).
\label{eq:blueline}
\end{equation}

Furthermore, in \cref{fig:comparison}, we also compared the performance of the Bell correlation depth witnesses from \cite{Baccari2018} with that of our DIWEDs. Since in \cite{Baccari2018} one takes the Bell inequalities of the form of \cref{eq:ineq6} and derives a nonlocality depth based on the assumption that parties in the same group can produce any nonsignalling (hence, potentially supraquantum) probability distribution, here we obtain tighter entanglement depth bounds because we restrict the power of the parties in the same group to produce only quantum correlations.

\section{Conclusions and outlook} \label{sec:concl}

In this work we have extended the analysis and results of \cite{Aloy2018}, presenting in detail the optimization procedure to derive DIWEDs from two-body permutationally invariant Bell inequalities. We have compared our findings to other existing criteria for determining entanglement depth and our methods yield better results (tighter bounds) when compared to methods based on stronger assumptions \cite{Baccari2018}, yet worse than methods that are not device-independent \cite{SorensenPRL2001, WinelandPRA1994}.

We see that a key ingredient in our method is the use of Jordan's lemma, which applies to Bell inequalities with two binary observables per site. Therefore, it is not obvious how to extend our method to some more general families of Bell inequalities than the ones we here consider, for instance, with more measurement settings per site \cite{WagnerPRL2017}. A natural next step, however, is the case of Bell inequalities with the same features, and with higher-order (but constant) correlators, which can be obtained without having to resort to the polytope approach \cite{FadelPRL2017}. In this case we would expect to obtain even tighter DIWEDs with essentially an analogous analysis as the one we performed here.

Another interesting feature we observe is that the optimal state tends to be symmetric at each partition. Recently, tremendous experimental progress has been achieved in measuring entanglement across different wells of a spin-squeezed BEC \cite{FadelScience2018, KunkelScience2018, LangeScience2018} (see also \cite{Shin2018}), which poses the question about how well can the DIWEDs we here present certify the entanglement in these systems, or how robust are the DIWED bounds for more realistic experimental conditions, such as finite temperature \cite{FadelQuantum2018}.

Throughout this work, our method to produce a lower bound certificate via semi-definite programming is currently based on the PPT condition as a relaxation of the separability condition. However, we expect that the gap between the variational solution and the certificate should not close in the general case, as fulfilling the PPT criterion is in general not sufficient to guarantee separability, even in the symmetric case \cite{TuraPRA2012, AugusiakPRA2012}, or even under stronger assumptions \cite{QuesadaPRA2014, QuesadaPRA2017, TuraQuantum2018}. Therefore, a natural next step in order to tighten these lower bound certificates is the use of stronger entanglement criteria (either using the symmetry condition in PPT states or using stronger, albeit computationally harder relaxations in the form of semidefinite programs \cite{DohertyPRA2004}).

\section*{Acknowledgments.}
This project has received funding from the European Union's Horizon 2020 research and innovation programme under the Marie-Sk{\l}odowska-Curie grant agreement No 748549,
Spanish MINECO (FISICATEAMO FIS2016-79508-P, QIBEQI FIS2016-80773-P, Severo Ochoa SEV-2015-0522, Severo Ochoa PhD fellowship), Fundacio Cellex, Generalitat de Catalunya (SGR 1341, SGR 1381 and CERCA/Program), ERC AdG OSYRIS and CoG QITBOX, EU FETPRO QUIC, the AXA Chair in Quantum Information Science. We thank M. Fadel for enlightening discussions regarding the experimental part. J. T. acknowledges support from   the   Alexander   von   Humboldt   Foundation. R.~A.~acknowledges the support from the Foundation for Polish Science through the First Team project (First TEAM/2017-4/31) co-financed by the European Union under the European Regional Development Fund.

\vfill

\bibliography{mylib}

\onecolumngrid
\appendix

\subsection{Asymptotic analysis details} \label{app:asymptotic}

In this section we motivate where the $\sigma = \sqrt{n/48}$ approximation, which we use in \cref{sec:asymptotic}, comes from. In \cite{AnnPhys} an asymptotic analysis of the value of the maximal quantum violation of \cref{eq:ineq6} was performed. However, only the scaling was relevant for that study. Here we determine the coefficient of the second order term exactly.

Let us recall that the expectation value of the optimal state (cf. \cref{eq:DickeSuperposition}) violating \cref{eq:ineq6} can be expressed as \cite{AnnPhys}:
\begin{equation}
\langle {\pazocal B}\rangle = \left(\frac{\beta_c}{n} - \frac{B}{2} + e^{-1/8\sigma}A'\right)n + (2B\sigma - A^2/2B + e^{-1/8\sigma}A') + O(\sigma/n),
\end{equation}
where $A = 2 \cos \theta$, $B = 4 \cos^2 \theta$ and $A' = -2 \sin \theta$, with $\theta = 5 \pi /6$ are the optimal asymptotic values. To determine de width $\sigma$ of the Gaussian superposition of Dicke states we use these values and we find that $\sigma$ must fulfill
\begin{equation}
e^{-1/8\sigma}(n+1) = 48 \sigma^2,
\end{equation}
which is a transcendental equation. However, since the scaling for $\sigma$ will be $\sqrt{n}$ in the first order, we can for large values of $n$ ignore the $e^{-1/8\sigma}$ term and therefore approximate $\sigma = \sqrt{(n+1)/48} \approx \sqrt{n/48}$, which is what we use in our asymptotic analysis.

\end{document}